\documentclass[aps,twocolumn,pra,tightenlines,floatfix,superscriptaddress]{revtex4-2}

\usepackage[breaklinks=true]{hyperref}
\usepackage{graphicx}
\usepackage[english]{babel}
\usepackage{amsmath}
\usepackage{amssymb}
\usepackage{times}

\usepackage{color}

\def\d{\mbox{d}}

\newcommand{\Ek}{{E^{}_{\mathbf{k}}}}

\newcommand{\sumk}{{\sum_{\mathbf{k}}}}

\newcommand{\uk}{u_{\mathbf{k}}}
\newcommand{\vk}{v_{\mathbf{k}}}

\begin{document}

\title{Rf spectra and pseudogap in ultracold Fermi gases across the BCS‑BEC crossover from pairing fluctuation theory}

 \author{Chuping Li}
 \affiliation{Hefei National Research Center for Physical Sciences at the Microscale and School of Physical Sciences, University 
   of Science and Technology of China, Hefei, Anhui 230026, China}
 \affiliation{Shanghai Research Center for Quantum Science and CAS Center for Excellence in Quantum Information and Quantum Physics, 
 University of Science and Technology of China, Shanghai 201315, China}
 \affiliation{Hefei National Laboratory, University of
  Science and Technology of China, Hefei 230088, China}
 \author{Lin Sun}
 \affiliation{Hefei National Laboratory, University of
  Science and Technology of China, Hefei 230088, China}
 \affiliation{Shanghai Research Center for Quantum Science and CAS Center for Excellence in Quantum Information and Quantum Physics, 
 University of Science and Technology of China, Shanghai 201315, China}
 \author{Kaichao Zhang}
 \affiliation{Hefei National Research Center for Physical Sciences at the Microscale and School of Physical Sciences, University 
   of Science and Technology of China, Hefei, Anhui 230026, China}
 \affiliation{Shanghai Research Center for Quantum Science and CAS Center for Excellence in Quantum Information and Quantum Physics, 
 University of Science and Technology of China, Shanghai 201315, China}
 \affiliation{Hefei National Laboratory, University of
  Science and Technology of China, Hefei 230088, China}
 \author{Junru Wu}
 \affiliation{Hefei National Research Center for Physical Sciences at the Microscale and School of Physical Sciences, University 
   of Science and Technology of China, Hefei, Anhui 230026, China}
 \affiliation{Shanghai Research Center for Quantum Science and CAS Center for Excellence in Quantum Information and Quantum Physics, 
 University of Science and Technology of China, Shanghai 201315, China}
 \affiliation{Hefei National Laboratory, University of
  Science and Technology of China, Hefei 230088, China}
 \author{Yuxuan Wu}
 \affiliation{Hefei National Research Center for Physical Sciences at the Microscale and School of Physical Sciences, University 
   of Science and Technology of China, Hefei, Anhui 230026, China}
 \affiliation{Shanghai Research Center for Quantum Science and CAS Center for Excellence in Quantum Information and Quantum Physics, 
 University of Science and Technology of China, Shanghai 201315, China}
 \affiliation{Hefei National Laboratory, University of
  Science and Technology of China, Hefei 230088, China}
 \author{Dingli Yuan}
 \affiliation{Hefei National Research Center for Physical Sciences at the Microscale and School of Physical Sciences, University 
   of Science and Technology of China, Hefei, Anhui 230026, China}
 \affiliation{Shanghai Research Center for Quantum Science and CAS Center for Excellence in Quantum Information and Quantum Physics, 
 University of Science and Technology of China, Shanghai 201315, China}
 \affiliation{Hefei National Laboratory, University of
  Science and Technology of China, Hefei 230088, China}
 \author{Pengyi Chen}
 \affiliation{Hefei National Research Center for Physical Sciences at the Microscale and School of Physical Sciences, University 
   of Science and Technology of China, Hefei, Anhui 230026, China}
 \affiliation{Shanghai Research Center for Quantum Science and CAS Center for Excellence in Quantum Information and Quantum Physics, 
 University of Science and Technology of China, Shanghai 201315, China}
 \affiliation{Hefei National Laboratory, University of
  Science and Technology of China, Hefei 230088, China}
 \author{Qijin Chen}
 \email[Corresponding author: ]{qjc@ustc.edu.cn}
 \affiliation{Hefei National Research Center for Physical Sciences at the Microscale and School of Physical Sciences, University 
   of Science and Technology of China, Hefei, Anhui 230026, China}
 \affiliation{Shanghai Research Center for Quantum Science and CAS Center for Excellence in Quantum Information and Quantum Physics, 
 University of Science and Technology of China, Shanghai 201315, China}
 \affiliation{Hefei National Laboratory, University of
  Science and Technology of China, Hefei 230088, China}

\date{\today}

\begin{abstract}

  The pseudogap phenomenon is a hallmark of strongly interacting Fermi
  systems, from high-temperature superconductors to ultracold atomic
  gases, yet its precise origin remains debated.  Here we calculate
  the spectral function and rf spectra of ultracold atomic gases
  across the BCS-BEC crossover to quantitatively investigate the
  pairing mechanism of the pseudogap. We advance our pairing
  fluctuation theory by incorporating particle-hole fluctuations,
  which renormalize the effective interaction in the particle-particle
  channel. To achieve quantitative accuracy, we employ a full
  numerical convolution for the pair susceptibility and self-energy,
  moving beyond previous analytic pseudogap approximations. This
  convolution approach automatically captures two critical effects:
  (i) the full spectral broadening of fermions due to finite pair
  lifetime, and (ii) the previously neglected pair-hole scattering
  effect, which manifests as a substantial Hartree energy. We
  calculate the spectral function, and use rf spectral intensity maps
  and energy distribution curves to determine the quasiparticle
  dispersion. From these, we extract the pseudogap $\Delta$, Hartree
  energy, and chemical potential, mapping their evolution across the
  crossover. Our results show that the pseudogap emerges continuously
  as the system moves from the BCS regime toward BEC. Furthermore, the
  pair spectral function reveals that pairs become diffusive at
  energies above $2\Delta$, indicating that the pair lifetime is
  governed by virtual binding and unbinding processes.  Our
  calculations achieve quantitative agreement with recent experiments
  across the BCS‑BEC crossover, including at unitarity, providing
  strong support for a pairing‑based origin of the pseudogap as
  described by our pairing fluctuation theory.

\end{abstract}

\maketitle

\section{INTRODUCTION}
\label{sec:1}

High-$T_{\text{c}}$ superconductors exhibit a distinctive feature compared to conventional superconductors: a gap in fermionic excitations in the normal state above $T_{\text{c}}$, known as the pseudogap \cite{Ding1996N,Timusk1999RPP,Leggett2006NP}. This phenomenon has sparked intense debate regarding its origin and interpretation. There is no consensus on whether the pseudogap originates from pairing or from competing orders, such as $d$-density wave \cite{DDW} or loop current orders \cite{Varma2014}.
Ultracold atomic Fermi gases provide a powerful quantum simulator for studying pseudogap phenomena, offering multiple experimentally tunable parameters \cite{Chen2005PR,Torma2016PC,Vale2021NP,Bloch2012NP}. In particular, the interaction strength can be tuned via an $s$-wave Feshbach resonance \cite{Chin2010RMP} from the weak-coupling Bardeen-Cooper-Schrieffer (BCS) limit to the strong-pairing Bose-Einstein condensation (BEC) limit, enabling exploration of the BCS-BEC crossover \cite{Leggett1980,Regal2005,Chen2005PR,Randeria2014ARCMP}.

To probe interaction effects, especially the pairing gap, radio-frequency (rf) spectroscopy has been used to measure spectral and thermodynamic quantities in ultracold Fermi gases, including the pairing gap \cite{Chin2004S,Perali2008PRL,Schirotzek140403PRL} and Tan's contact \cite{Stewart2010PRL,Sagi2012PRL,Mukherjee2019PRL}. With momentum resolution, this technique has been used to measure the occupied spectral intensity, from which the fermion spectral function $A(\mathbf{k},\omega)$ can be extracted \cite{Stewart2008N,Gaebler2010NP,Feld2011N,Sagi2015PRL}. The spectral function represents the probability density for a single-particle excitation with momentum $\mathbf{k}$ and energy $\omega$, providing direct evidence of a pseudogap in strongly interacting Fermi gases \cite{Chen2009PRL,Magierski2009PRL}. A recent experiment on homogeneous $^6$Li gases using  momentum-resolved microwave spectroscopy has clarified the existence and origin of the pseudogap, revealing that it arises from pair fluctuations \cite{Li2024N}.
Accurate theoretical models that capture both qualitative and quantitative aspects are essential for further experimental progress across the entire BCS-BEC crossover.

In this work, we employ a pairing fluctuation theory that incorporates
contributions from the particle-hole channel to provide an enhanced
description of strongly interacting Fermi systems. We develop an
iterative framework to calculate the spectral function, expressing the
pair susceptibility and fermion self-energy as convolutions in the
real-frequency domain. This framework enables the computation of the
average Hartree energy, the physical chemical potential, the fermion
and pair spectral functions, and the single-particle density of states
(DOS). Using this numerical approach, we analyze the evolution of
spectral functions near the superfluid transition temperature across
the BCS-BEC crossover. Our results show that as the interaction
strength increases from the BCS to the BEC regime, the pairing gap
grows monotonically, as evidenced by the DOS and energy distribution
curves (EDCs) derived from the fermion spectral functions. We compare
the pairing gap extracted from EDCs and simulated rf spectra with
experimental data and find quantitative agreement. Additionally, we
compute the spectrum of the two-particle propagator and use it to
explain the excitation behavior of finite-momentum pairs. A focus on
the unitary limit and quantitative comparison of the temperature
dependence of the extracted pairing gap with recent momentum-resolved
microwave spectroscopy data \cite{Li2024N} is given in a companion
paper \cite{letter}.

This paper is organized as follows.  In Sec.~\ref{sec:2}, we present
an overview of the pairing fluctuation theory, including contributions
from the particle-hole channel, and introduce an iterative framework for
calculating spectral functions.  In Sec.~\ref{sec:3}, we present
numerical results for the average Hartree energy, the physical
chemical potential, spectral functions, and the DOS across the BCS-BEC
crossover, with comparisons to experimental data.  Finally, we
summarize in Sec.~\ref{sec:4} our findings and discuss  their
implications.

\section{THEORETICAL FORMALISM}
\label{sec:2}

\subsection{Pairing Fluctuation Theory without Particle-hole Fluctuations}
\label{sec:2a}

In this section, we first present an overview of the pairing fluctuation theory \cite{Chen1998PRL,Chen1999PRB,Chen2005PR} \emph{without} the particle-hole channel contributions, and then incorporate this contribution as the foundation for calculating spectral functions using an iterative framework. We consider a Hamiltonian describing a homogeneous three-dimensional (3D) Fermi gas,
\begin{equation}
  \label{eq:Hamiltonian}
  H =\sum_{\mathbf{k} \sigma} \epsilon_{\mathbf{k}} c_{\mathbf{k} \sigma}^{\dagger} c_{\mathbf{k} \sigma} 
  \!+\!g\!\sum_{\mathbf{k} \mathbf{k}' \mathbf{q}} c_{\mathbf{k}+\frac{\mathbf{q}}{2} \uparrow}^{\dagger} c_{-\mathbf{k}+\frac{\mathbf{q}}{2} \downarrow}^{\dagger} c_{-\mathbf{k}'+\frac{\mathbf{q}}{2} \downarrow} c_{\mathbf{k}'+\frac{\mathbf{q}}{2} \uparrow}\,,
\end{equation}
where $\epsilon_{\mathbf{k}} = \mathbf{k}^2/2m$ and $g < 0$ is a short-range $s$-wave attractive interaction. We have the Fermi momentum $k_\text{F}=(3\pi^{2}n)^{1/3}$ and the Fermi energy $E_\text{F} \equiv k_\text{B}T_\text{F} = \hbar^{2}k_\text{F}^{2}/2m$, where $n$ is the number density and $m$ is the atomic mass. Four-momenta are denoted as $K \equiv (\mathbf{k},\mathrm{i}\omega_n)$ and $Q \equiv (\mathbf{q},\mathrm{i}\Omega_l)$, with $\sum_K \equiv T\sum_n\sum_\mathbf{k}$, where $\omega_n$ and $\Omega_l$ are odd and even Matsubara frequencies, respectively, following Ref.~\cite{Chen1998PRL}. Throughout, we use the natural units $\hbar=k_\text{B}=1$ and set the volume to unity.

We adopt the pairing fluctuation theory  developed for the pseudogap physics in the cuprates \cite{Chen1998PRL}, which has been extended to the BCS-BEC crossover in ultracold Fermi gases \cite{Chen2005PR}. The full $T$-matrix is composed of contributions from both zero-momentum pairs $t_\text{sc}(Q)$ and nonzero-momentum pairs $t_\text{pg}(Q)$,
\begin{subequations}
\begin{eqnarray}
  t(Q) &=& t_\text{sc}(Q) + t_\text{pg}(Q)\,,\\
  t_\text{sc}(Q)&=&-\frac{\Delta_\text{sc}^2}{T}\delta(Q)\,,\\
  t_\text{pg}(Q)&=&\frac{g}{1+g \chi(Q)}\,,
\end{eqnarray}
\end{subequations}
with the superfluid order parameter $\Delta_\text{sc}$ and the pair
susceptibility $\chi(Q) = \sum_K G_0(Q-K)G(K)$, where
$G_0(K)=(\mathrm{i}\omega_n-\xi_{\mathbf{k}})^{-1}$ and $G(K)$
represent the non-interacting and full Green's functions,
respectively, with the free fermion dispersion $\xi_{\mathbf{k}} =
\epsilon_{\mathbf{k}} - \mu'$ and the shifted chemical potential
$\mu'$. The superconducting $t_\text{sc}(Q)$ vanishes above
$T_{\text{c}}$, while the pseudogap $t_\text{pg}(Q)$, a two-particle
propagator, remains present both above and below $T_{\text{c}}$.

The full self-energy $\Sigma(K)$ thus comprises contributions from the Cooper pair condensate $\Sigma_\text{sc}(K)$ and finite-momentum pairs $\Sigma_\text{pg}(K)$,
\begin{subequations}
\begin{eqnarray}
  \Sigma(K) &=& \Sigma_\text{sc}(K) + \Sigma_\text{pg}(K)\,,\\
  \Sigma_\text{sc}(K) &=& -\Delta_\text{sc}^2\, G_0(-K)\,,\\
  \label{eq:sigmapg}
  \Sigma_\text{pg}(K) &=& \sum_Q t_\text{pg}(Q)\, G_0(Q-K)\,.
\end{eqnarray}
\end{subequations}

The Thouless criterion, which determines the superfluid transition, requires $t_\text{pg}^{-1}(0) = g^{-1} + \chi(0) = 0$ at $T_{\text{c}}$, implying that $t_\text{pg}(Q)$ peaks strongly at $Q = 0$. Thus, the dominant contribution to $\Sigma_\text{pg}(K)$ arises near $Q = 0$, allowing the approximation
\begin{equation}
  \label{eq:Sigmapgappro}
  \Sigma_\text{pg}(K) \approx \biggl[\sum_Q t_\text{pg}(Q)\biggr] G_0(-K) = -\Delta_\text{pg}^2 G_0(-K)\,,
\end{equation}
where the pseudogap is defined as $\Delta_\text{pg}^2=-\sum_{Q\neq0}t_\text{pg}(Q)$. 
Thus, without heavy numerics, we arrive at a total gap $\Delta^2 = \Delta_\text{sc}^2 + \Delta_\text{pg}^2$, which appears in a BCS-like self-energy, 
\begin{equation}
  \label{eq:Sigma}
  \Sigma(K) \approx \frac{\Delta^2}{\mathrm{i}\omega_n + \xi_\mathbf{k}}\,.
\end{equation}
Note that the Hartree energy has been absorbed into $\mu'$ (hence into
$G_0(K)$ and $\Sigma(K)$) through the equation-of-motion approach
\cite{Kadanoff1961,ChenPhD}, which is required to satisfy $\mu'= 0$ on
the Fermi surface. (In this way, when the ``off-diagonal'' gap
vanishes, only the Hartree energy is present as a shift to the
chemical potential.)  Then the full Green's function takes the BCS
form,
\begin{equation}
  \label{eq:Gfunc}
  G(K)=\frac{u_{\mathbf{k}}^{2}}{\mathrm{i}\omega_{n}-E_{\mathbf{k}}}+\frac{v_{\mathbf{k}}^{2}}{\mathrm{i}\omega_{n}+E_{\mathbf{k}}}\,,
\end{equation}
with the coherence factors
$u_{\mathbf{k}}^{2}=(1+\xi_{\mathbf{k}}/E_{\mathbf{k}})/2$ and
$v_{\mathbf{k}}^{2}=(1-\xi_{\mathbf{k}}/E_{\mathbf{k}})/2$, along with
the Bogoliubov quasiparticle dispersion
$E_{\mathbf{k}}=\sqrt{\xi_{\mathbf{k}}^{2}+\Delta^{2}}$. Then
 the  constraint $n = 2\sum_K G(K)$ yields the
number equation
\begin{equation}
  \label{eq:num}
  n=2\sum_{\mathbf{k}}\left[v^2_{\mathbf{k}}+\frac{\xi_{\mathbf{k}}}{E_{\mathbf{k}}}f(E_{\mathbf{k}})\right]\,,
\end{equation}
where $f(x)$ is the Fermi distribution function. 

With the known form of $G(K)$, the inverse $T$-matrix $t_\text{pg}^{-1}(Q)$ can be Taylor expanded on the real-frequency axis. 
After analytic continuation, we have 
$t_\text{pg}^{-1}(\mathbf{q},\Omega) \approx a_\text{1}\Omega^{2}+a_\text{0}(\Omega-\Omega_{\mathbf{q}}+\mu_\text{p})$, 
where $\Omega_{\mathbf{q}}={\mathbf{q}}^{2}/2M$, with $M$  the effective pair mass. 
The coefficients $a_{0}$, $a_{1}$ and $1/2M$ are obtained during the expansion. 
The definition of $\Delta^2_\text{pg}$ then leads to the pseudogap equation
\begin{equation}
  |a_0|\Delta_\text{pg}^{2}=\sum_{\mathbf{q}}\frac{b(\tilde{\Omega}_{\mathbf{q}})}{\sqrt{1+4\dfrac{a_{1}}{a_0}(\Omega_{\mathbf{q}}-\mu_\text{p})}}\,,
  \label{eq:pg}
\end{equation}
where $b(x)$ is the Bose distribution function, with the pair dispersion 
$\tilde{\Omega}_{\mathbf{q}} = \bigl[\sqrt{a_0^{2} + 4 a_1 a_0 (\Omega_{\mathbf{q}} - \mu_\text{p})} - a_0\bigr]/{2 a_1}.$
In the BEC regime, where $a_1/a_0$ is small, the $a_1$ term provides a minor correction, and $\tilde{\Omega}_{\mathbf{q}} \approx \Omega_{\mathbf{q}} - \mu_\text{p}$, where $\mu_\text{p}$ is the pair chemical potential.

This expansion yields  $g^{-1}+\chi(0) = a_0 \mu_\text{p}$, which must necessarily vanish for $T \le T_\text{c}$ a la the Thouless criterion.
Except at temperatures far above $T_c$, we have an extended BCS gap equation,
\begin{equation}
  \label{eq:gap}
  a_0\mu_\text{p}=\frac{m}{4\pi a}+\sum_{\mathbf{k}}\left[\frac{1-2f(E_{\mathbf{k}})}{2E_{\mathbf{k}}}-\frac{1}{2\epsilon_\mathbf{k}}\right]\,,
\end{equation}
which has been regularized using the Lippmann-Schwinger relation $g^{-1}={m}/{4\pi a}-\sum_{\mathbf{k}}{1}/{2\epsilon_\mathbf{k}}$, 
with $a$ the $s$-wave scattering length. Note that, far above $T_c$, the pseudogap approximation given in Eq.~\eqref{eq:Sigmapgappro} is no longer valid, so that Eq.~\eqref{eq:gap} will also break down. Nonetheless, its solution may serve as a good initial input for numerical iterative procedures.

\subsection{Effects of the Particle-hole Channel}
\label{sec:2b}

In ultracold Fermi gases, superfluidity across the BCS-BEC crossover mainly involves pairing in the particle-particle channel. The particle-hole channel, often referred to as induced interactions in the literature \cite{GORKOV1961}, introduces corrections to the pair susceptibility, which significantly affect the crossover regime by reducing the effective interaction strength and shifting the $T_{\text{c}}$ curve toward the BEC regime \cite{Chen2016SR}. To improve agreement with experiment, we take into account the contributions from the particle-hole channel. Following Ref.~\cite{Chen2016SR}, we include a particle-hole susceptibility $\chi_{\text{ph}}(P) = \sum_K G(K) G_0(K-P)$, which self-consistently incorporates the self-energy feedback, where $P = (\mathbf{p},\mathrm{i}\nu_n)$ denotes the total four-momentum of the particle-hole propagator. The resulting full $T$-matrix, combining both particle-particle and particle-hole contributions, satisfies $t_\text{pg}^{-1}(Q) = g^{-1} + \chi_{\text{ph}}(P) + \chi(Q)$.

To get rid of the complicated dependence on external momenta introduced through $\chi_{\text{ph}}(P)$,  an average of $\chi_{\text{ph}}(P)$ is taken on the Fermi surface for on-shell elastic scattering, a method commonly used in the literature when studying induced interactions \cite{GORKOV1961,Yin2009}, as fermions near the Fermi surface dominate the pairing channel. Following Ref.~\cite{Chen2016SR},
we set $\mathrm{i}\nu_n = 0$, and obtain 
\begin{eqnarray}
\chi_{\text{ph}}(\mathbf{p},0) \!\!
&= {\displaystyle \sum_\mathbf{k}}\!\! &\left[ \frac{f(\Ek)-f(\xi_\mathbf{k-p})}
  {\Ek-\xi_\mathbf{k-p}} \uk^2 \right.\nonumber\\
&& {}\left.  -\frac{1-f(\Ek)-f(\xi_\mathbf{k-p})}
  {\Ek+\xi_\mathbf{k-p}}\vk^2 \right]\,.
\label{eq:chi_ph}
\end{eqnarray}
Here $\mathbf{p}$ is restricted to $|\mathbf{p}| = |\mathbf{k} +
\mathbf{k}'| = k_{\mu} \sqrt{2(1+\cos\theta)}$, where $|\mathbf{k}| =
|\mathbf{k}'| = k_{\mu}$ are the momenta of on-shell elastic
scattering, and $\theta$ is the angle between $\mathbf{k}$ and
$\mathbf{k}'$. 
Averaging Eq.~\eqref{eq:chi_ph}  over
the angle $\theta$ yields $\langle\chi_{\text{ph}}\rangle $ and
 simplifies $t_\text{pg}(Q)$ to
\begin{equation}
  t_\text{pg}(Q) = \frac{1}{g^{-1} + \langle\chi_{\text{ph}}\rangle + \chi(Q)}\,.
\end{equation}
Thus, the gap equation, incorporating the particle-hole contribution, becomes
\begin{equation}
  \label{eq:gapph}
  a_0\mu_\text{p} = \langle\chi_{\text{ph}}\rangle + \frac{m}{4\pi a} + \sum_{\mathbf{k}}\left[\frac{1-2f(E_{\mathbf{k}})}{2E_{\mathbf{k}}} - \frac{1}{2\epsilon_\mathbf{k}}\right]\,,
\end{equation}
while the other equations remain unchanged. It is known that $\langle\chi_{\text{ph}}\rangle <0$ so that it provides a screening to the bare pairing interaction, leading to a reduced effective pairing strength. 

Finally, Eqs.~\eqref{eq:num}, \eqref{eq:pg}, and \eqref{eq:gapph}, along with the average of Eq.~\eqref{eq:chi_ph}, form a closed set of self-consistent equations, which can be used to solve for $(\mu', \Delta, \mu_\text{p})$ at $T > T_{\text{c}}$, for $(\mu', \Delta_\text{pg}, T_{\text{c}})$ with $\Delta_\text{sc}=0$, and for $(\mu', \Delta, \Delta_\text{pg})$ at $T < T_{\text{c}}$. Here the order parameter $\Delta_\text{sc}$ can be derived from $\Delta_\text{sc}^2 = \Delta^2 - \Delta_\text{pg}^2$ below $T_{\text{c}}$. These solutions may serve as the initial input parameters for subsequent calculations of spectral functions.

\subsection{Iterative Framework for Computing the Spectral Functions}

\begin{figure}
\centering
\includegraphics[clip,width=3.4in]{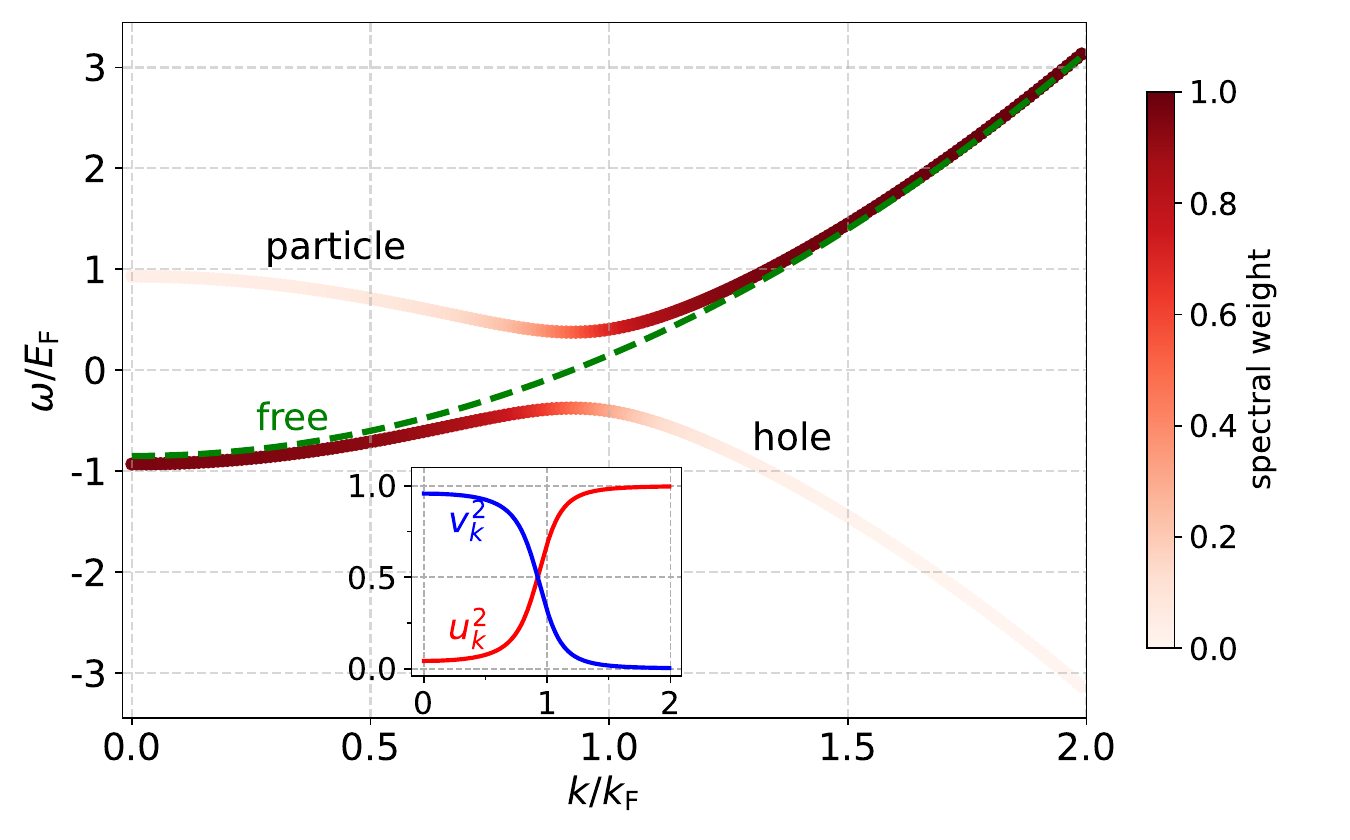}
\caption{$A_\text{ini}(\mathbf{k},\omega)$ at $T_{\text{c}}$ and
  unitarity.  The upper and lower curves represent particle and hole
  branches of the dispersion, respectively, color-coded using the
  spectral weight given by the coherence factors $u_{\mathbf{k}}^{2}$
  and $v_{\mathbf{k}}^{2}$, shown in the inset as a function of
  $k/k_\text{F}$. For comparison, also plotted as the green dashed
  curve is the free fermion dispersion.  }
\label{fig:Akwinitial}
\end{figure}

Using the BCS-like Green's function in Eq.~\eqref{eq:Gfunc}, the initial fermion spectral function $A_\text{ini}(\mathbf{k},\omega)$ is given by 
\begin{equation}
  \label{eq:iniAkw}
  A_\text{ini}(\mathbf{k},\omega) = 2\pi\bigl[u_\mathbf{k}^{2}\,\delta(\omega-E_\mathbf{k}) + v_\mathbf{k}^{2}\,\delta(\omega+E_\mathbf{k})\bigr]\,. 
\end{equation}
We present in Fig.~\ref{fig:Akwinitial} a typical spectral function, $A_\text{ini}(\mathbf{k},\omega)$, calculated at $T_{\text{c}}$ and unitarity ($1/k_\text{F}a = 0$), with $\Delta/E_{\text{F}}=0.376$ and $\mu'/E_{\text{F}}=0.85$. There exist two distinct branches, representing particle and hole quasiparticle excitations, as labeled. The color coding represents the spectral weight, given by the coherence factors shown in the inset. Both branches exhibit a back-bending behavior near $k \approx 0.93k_\text{F}$, a characteristic feature of BCS-like dispersions. The green dashed curve shows the free fermion dispersion for reference.
Under the pseudogap approximation given by
Eq.~\eqref{eq:Sigmapgappro}, the BCS form of the Green's function
neglects any broadening effect; consequently,
$A_\text{ini}(\mathbf{k},\omega)$ cannot provide useful EDCs or pair
lifetime information, and thus cannot produce realistic rf spectra.
This necessitates an iterative framework beyond the pseudogap
approximation in order to capture these effects and compare with
experimental data.

To address the limitations of $A_\text{ini}(\mathbf{k},\omega)$, such as the absence of spectral broadening, we propose an iterative framework that directly evaluates the convolution in Eq.~\eqref{eq:sigmapg} on the real-frequency axis, thereby bypassing the approximation in Eq.~\eqref{eq:Sigmapgappro}.

We proceed with spectral representations, where the full Green's
function is expressed as
\begin{equation}
  G({\mathbf{k}},\mathrm{i}\omega_{n}) = -\int\frac{{\mathrm{d}}\omega}{\pi}\frac{\text{Im}\, G^{\text{R}}({\mathbf{k}},\omega)}{\mathrm{i}\omega_{n}-\omega}\,.
\end{equation}
Substituting it into the pair susceptibility $\chi(Q)$ and performing Matsubara sums yields the retarded $\chi^{\text{R}}({\mathbf{q}},\Omega)$, 
\begin{subequations}
\label{eq:susconv}
\begin{eqnarray}
  \text{Re}\chi^{\text{R}}({\mathbf{q}},\Omega) &=& \sum_{\mathbf{k}}\int\frac{{\mathrm{d}}\omega}{\pi}\frac{1-f(\omega)-f(\xi_{{\mathbf{q}}-{\mathbf{k}}})}{\Omega-\omega-\xi_{{\mathbf{q}}-{\mathbf{k}}}}\nonumber\\
  &&{}\times \text{Im}\,G^{\text{R}}({\mathbf{k}},\omega)\,,\\
  \label{eq:susconva}
  \text{Im}\chi^{\text{R}}({\mathbf{q}},\Omega) &=& -\sum_{\mathbf{k}}\bigl[1-f(\xi_{{\mathbf{q}}-{\mathbf{k}}})-f(\Omega-\xi_{{\mathbf{q}}-{\mathbf{k}}})\bigr]\nonumber\\
  &&{}\times \text{Im}\,G^{\text{R}}({\mathbf{k}},\Omega-\xi_{{\mathbf{q}}-{\mathbf{k}}})\,.
  \label{eq:susconvb}
  \end{eqnarray}
\end{subequations}
This leads to
\begin{equation}
  \label{eq:tpgconv}
  \text{Im}\,t_\text{pg}^{\text{R}}({\mathbf{q}},\Omega)\!=\!\frac{-\text{Im}\chi^{\text{R}}({\mathbf{q}},\Omega)}{[g^{-1}\!\!+\!\langle\chi_{\text{ph}}\rangle\!+\!\text{Re}\chi^{\text{R}}({\mathbf{q}},\Omega)]^{2}\!+\![\text{Im}\chi^{\text{R}}({\mathbf{q}},\Omega)]^2}\,,
\end{equation}
and
\begin{equation}
  t_{\text{pg}}({\mathbf{q}},\mathrm{i}\Omega_{l})=g-\int{\frac{{\mathrm{d}}\Omega}{\pi}}\frac{\text{Im}\,t_{\text{pg}}^{\text{R}}({\mathbf{q}},\Omega)}{\mathrm{i}\Omega_{l}-\Omega}\,.
\end{equation}
Note that $t({\mathbf{q}},\mathrm{i}\Omega_{l}\rightarrow \infty) = g $ rather than zero, so that it needs to be subtracted in the Kramers-Kronig relations.
Substituting into Eq.~\eqref{eq:sigmapg} yields 
\begin{subequations}
\label{eq:Sigmapgconv}
\begin{eqnarray}
  \text{Re}\Sigma^{\text{R}}_{\text{pg}}({\mathbf{k}},\omega)&=&-\sum_{\mathbf{q}}\int{\frac{{\mathrm{d}}\Omega}{\pi}}{\frac{b(\Omega)+f(\xi_{{\mathbf{q}}-{\mathbf{k}}})}{\xi_{{\mathbf{q}}-{\mathbf{k}}}-\Omega+\omega}}\nonumber\\&&{}\times{\text{Im}\,t^{\text{R}}_{\text{pg}}({\mathbf{q}},\Omega)}\,,\\
  \label{eq:Sigmapgconva}
  \text{Im}\Sigma^{\text{R}}_{\text{pg}}({\mathbf{k}},\omega)&=&\sum_{\mathbf{q}}\left[b(\xi_{{\mathbf{q}}-{\mathbf{k}}}+\omega)+f(\xi_{{\mathbf{q}}-{\mathbf{k}}})\right]\nonumber\\&&{}\times\text{Im}\,t^{\text{R}}_{\text{pg}}({\mathbf{q}},\xi_{{\mathbf{q}}-{\mathbf{k}}}+\omega)\,.
  \label{eq:Sigmapgconvb}
\end{eqnarray}
\end{subequations}
The expressions for $t_\text{sc}(Q)$ and $\Sigma_\text{sc}(K)$ remain unchanged. 
Then we have
\begin{equation}
  {\text{Im}}\,G^{\text{R}}({\mathbf{k}},\omega)\!=\!\frac{{\text{Im}}\,\Sigma^{\text{R}}({\mathbf{k}},\omega)}{[\omega\!-\!\epsilon_{\mathbf{k}}\!+\!\mu\!-\!{\text{Re}}\,\Sigma^{\text{R}}({\mathbf{k}},\omega)]^2\!+\![{\text{Im}}\,\Sigma^{\text{R}}({\mathbf{k}},\omega)]^2}\,,
  \label{eq:Gfuncconv}
\end{equation}
which leads to the broadened spectral function, 
\begin{equation}
  A({\mathbf{k}},\omega)=-2\,{\text{Im}}\,G^{\text{R}}({\mathbf{k}},\omega)\,.
  \label{eq:Akwn}
\end{equation}
Unlike $A_\text{ini}(\mathbf{k},\omega)$ derived from
Eq.~(\ref{eq:Gfunc}), $A({\mathbf{k}},\omega)$ includes quasiparticle
lifetime effects \cite{letter}.  Here $\mu = \mu' +
\bar{E}_{\text{Hartree}}$ is the physical chemical potential, as the
Hartree energy is absorbed into the self-energy.  We assume the
average Hartree energy $\bar{E}_{\text{Hartree}}$ is real and
$K$-independent, causing a chemical potential shift and minor mass
renormalization \cite{letter}.  As an incoherent background
contribution from the diagonal term of $\Sigma_\text{pg}(K)$,
$\bar{E}_{\text{Hartree}}$ is given by
$\bar{E}_{\text{Hartree}}=\text{Re}\,\Sigma^\text{R}_\text{pg}(k_\mu,0)$,
where $k_{\mu}=\sqrt{2m\mu'}$ is the wave vector on the Fermi surface.  In
the BEC regime, where $\mu' < 0$, $\bar{E}_{\text{Hartree}}$ is
extracted from the dispersion of the hole branch using Eq.~(29) of
Ref.~\cite{Chen2009RoPP}.

\begin{figure}
\centering
\includegraphics[clip,width=3.2in] {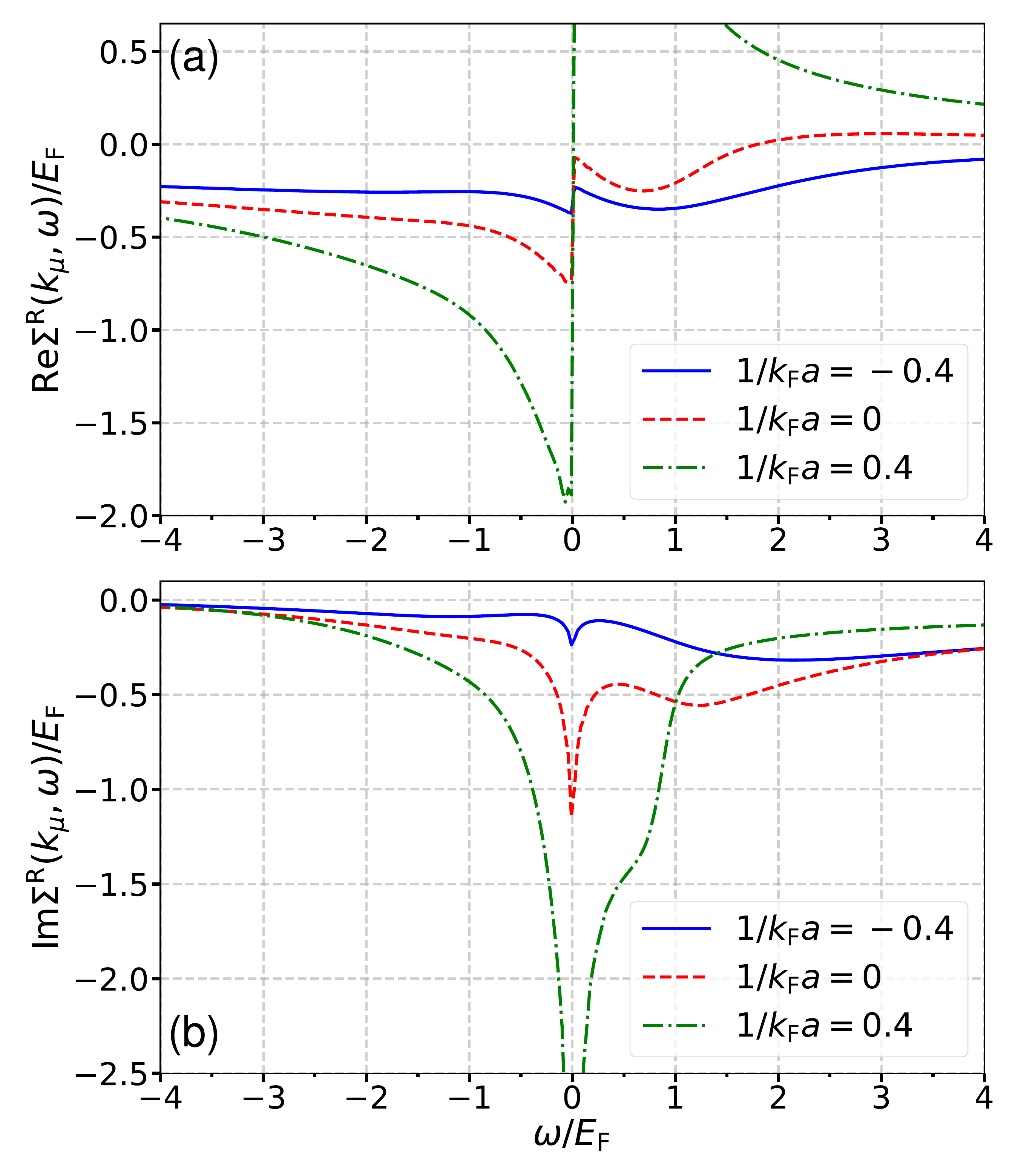}
\caption{ (a) Real and (b) imaginary part of the retarded self-energy
  $\Sigma^{\text{R}}(k_{\mu},\omega)$ of a homogeneous Fermi gas on
  the Fermi surface at $T_{\text{c}}$ for typical interactions across
  the BCS-BEC crossover.  The negative peak at $\omega=0$ in (b) is
  associated with the appearance of the pseudogap.}
\label{fig:sigma}
\end{figure}

In Fig.~\ref{fig:sigma}, we present the real and imaginary parts of
the retarded self-energy $\Sigma^\text{R}(\mathbf{k},\omega)$ at
$|\mathbf{k}|=k_{\mu}$ for a homogeneous Fermi gas at $T_\text{c}$ and
various $1/k_\text{F}a$, computed using Eq.~(\ref{eq:Sigmapgconv})
with $\Sigma_\text{sc}(K)=0$.  Unlike the simple BCS-like self-energy
given in Eq.~(\ref{eq:Sigma}), the real part
$\text{Re}\Sigma^\text{R}(k_\mu,\omega)$ is no longer antisymmetric
about the Fermi surface ($\omega=0$), due to the contribution from
finite-momentum pairing and fermion scattering.  According to
Eq.~(\ref{eq:Gfuncconv}), the solutions of the equation
$\omega+\bar{E}_{\text{Hartree}}=\text{Re}\Sigma^{\text
  R}(k_\mu,\omega)$ roughly correspond to the positions of
quasiparticle peaks in the EDCs, while the imaginary part
$\text{Im}\Sigma^\text{R}(k_\mu,\omega)$, which exhibits a negative
peak at the Fermi surface, determines the quasiparticle lifetime and
contributes to the pairing gap in the fermion spectral function. The
latter can be seen from the fact that, at the Fermi level, $A(k_\mu,0)
= -2/\text{Im}\,\Sigma^\text{R}(k_\mu,0)$ becomes a minimum at
$\omega=0$, hence leaving two side peaks in the spectral function. The
larger negative peak is in agreement with the increased pseudogap
parameter as the system evolves from BCS to BEC.

\begin{figure}
\centering
\includegraphics[clip,width=3.4in] {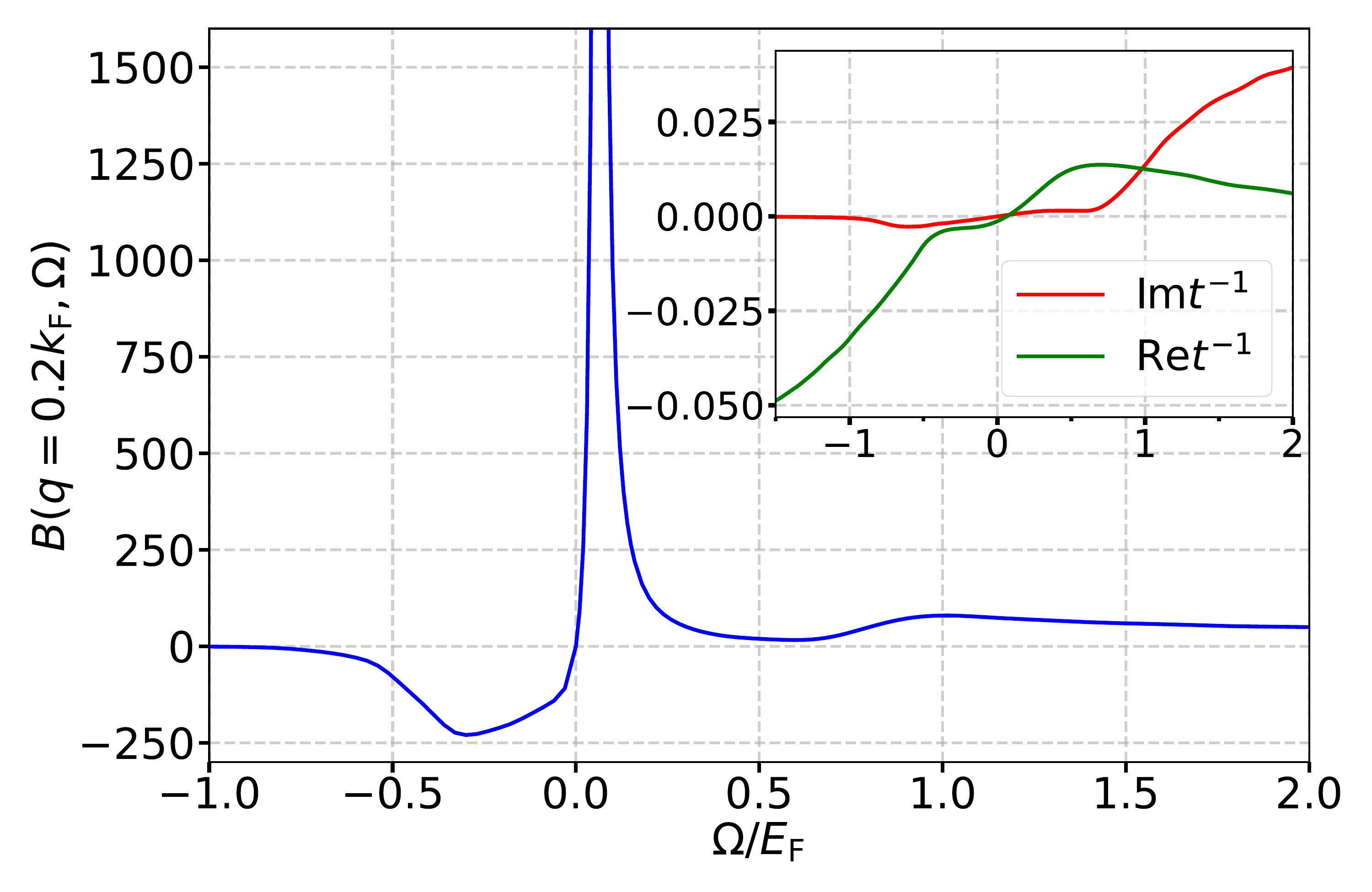}
\caption{ Pair spectral function $B({\mathbf{q}},\Omega)$ at
  $|\mathbf{q}|=0.2k_{\text{F}}$ and $T_{\text{c}}$ for
  $1/k_{\text{F}}a=0$.  Plotted in the inset are the real and imaginary
  parts of the corresponding inverse $T$-matrix $t^{-1}(\mathbf{q},
  \Omega)$, as labeled.  }
\label{fig:Tmatrix}
\end{figure}

Using $\text{Im}\,G^{\text{R}}({\mathbf{k}},\omega)$ obtained from
Eq.~(\ref{eq:Gfuncconv}), we recompute
$\chi^{\text{R}}({\mathbf{q}},\Omega)$ via Eq.~(\ref{eq:susconv}) and
$\text{Im}\,t_\text{pg}^{\text{R}}({\mathbf{q}},\Omega)$ via
Eq.~(\ref{eq:tpgconv}) to obtain the pair spectral function, which is
defined as
\begin{equation}
B({\mathbf{q}},\Omega)=-2\,\text{Im}\,t^{\text{R}}({\mathbf{q}},\Omega)\,,
\label{eq:Bkw}
\end{equation}
representing the likelihood of pair excitations at $(\mathbf{q},
\Omega)$.  Shown in Fig.~\ref{fig:Tmatrix} is a representative pair
spectral function $B({\mathbf{q}},\Omega)$ as a function of $\Omega$,
calculated for $|\mathbf{q}|=0.2k_{\text{F}}$ at $T_{\text{c}}$ and
$1/k_{\text{F}}a=0$.  The narrow peak near $\Omega=0$ indicates the
formation of long-lived finite-momentum pairs.  The tail at large
positive frequencies arises from an incoherent background, while the
contributions at negative frequencies represent the zero-point energy
resulting from quantum fluctuations.  Shown in the inset are the real
and imaginary parts of the inverse $T$-matrix $t^{-1}(\mathbf{q},
\Omega)$.  Here $\text{Re}\,t^{-1}(\mathbf{q}, \Omega)$ crosses zero
near $\Omega=0$, corresponding to the peak location in
$B({\mathbf{q}},\Omega)$.  The rapid increase in
$\text{Im}\,t^{-1}(\mathbf{q}, \Omega)$ at higher $\Omega \gtrsim
0.7E_F $ signals that pairs become diffusive and short-lived at energy
above this threshold, determined by the minimum energy of the
two-particle continuum.

For numerical calculations, we compute
$\Sigma^{\text{R}}_{\text{pg}}({\mathbf{k}},\omega)$ in
Eq.~(\ref{eq:Sigmapgconv}) using an adaptive quadrature method.  To
improve precision, we sample $\Omega$ at varying step sizes for given
$q=|\mathbf{q}|$, adding more points where
$\text{Im}\,t_\text{pg}^{\text{R}}({\mathbf{q}},\Omega)$ changes
rapidly.  For stable pairs near $Q = 0$ below $T_{\text{c}}$,
particularly in the BEC regime, the negative peak in 
$\text{Im}\,t_\text{pg}^{\text{R}}({\mathbf{q}},\Omega)$ may become extremely sharp so that it is difficult to integrate numerically, as shown in Fig.~\ref{fig:Tmatrix} (or even become a true delta function).  We then
approximate such a sharp peak with a delta function and treat it separately,
\begin{equation}
  \text{Im}\,t_\text{pg}^{\text{R}}({\mathbf{q}},\Omega) \approx \text{Im}\,t'({\mathbf{q}},\Omega)-\frac{\pi}{a_0}\delta(\Omega-\Omega_{\mathbf{q}}) \,,
\end{equation}
where $\text{Im}\, t'(\mathbf{q}, \Omega)$ is the imaginary part of
$t_\text{pg}^\text{R}({\mathbf{q}},\Omega)$ with the sharp peak
subtracted.

\begin{figure}
\centering
\includegraphics[clip,width=3.2in] {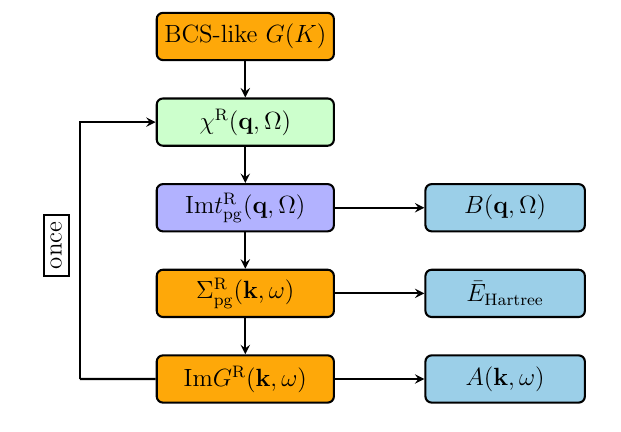}
\caption{Flowchart of the iterative framework for calculating
  $\bar{E}_{\text{Hartree}}$, $A({\mathbf{k}},\omega)$, and
  $B({\mathbf{q}},\Omega)$. }
\label{fig:flowchart}
\end{figure}

The flowchart in Fig.~\ref{fig:flowchart} outlines the iterative
framework for calculating the spectral functions
$A({\mathbf{k}},\omega)$ and $B({\mathbf{q}},\Omega)$, along with the
average Hartree energy $\bar{E}_{\text{Hartree}}$.  Using the
self-consistent solutions of Eqs.~(\ref{eq:num}), (\ref{eq:pg}), and
(\ref{eq:gapph}) as input, we initialize the process with the BCS-like
full Green's function from Eq.~(\ref{eq:Gfunc}) as the starting point
for computing $\chi^{\text{R}}({\mathbf{q}},\Omega)$ in
Eq.~(\ref{eq:susconv}).  We then compute
$\text{Im}\,t_\text{pg}^{\text{R}}({\mathbf{q}},\Omega)$ using
Eq.~(\ref{eq:tpgconv}), followed by
$\Sigma^{\text{R}}_{\text{pg}}({\mathbf{k}},\omega)$ and the
corresponding $\bar{E}_{\text{Hartree}}$ via
Eq.~(\ref{eq:Sigmapgconv}).  The resulting
${\text{Im}}\,G^{\text{R}}({\mathbf{k}},\omega)$ from
Eq.~(\ref{eq:Gfuncconv}) yields the broadened
$A({\mathbf{k}},\omega)$.  To obtain $B({\mathbf{q}},\Omega)$, we
iterate by recomputing $\chi^{\text{R}}({\mathbf{q}},\Omega)$ and
$\text{Im}\,t_\text{pg}^{\text{R}}({\mathbf{q}},\Omega)$ using the
updated ${\text{Im}}\,G^{\text{R}}({\mathbf{k}},\omega)$.  This
framework captures broadening effects that are absent in
$A_\text{ini}(\mathbf{k},\omega)$.

In practice, after the first iteration, the pair susceptibility
$\chi(Q)$ leads to a deviation from the Thouless criterion
$g^{-1}+\chi(0)=0$ below $T_{\text{c}}$.  We check the deviation,
$\tau \equiv g^{-1}+\chi(0)$, and find that it remains small, with
$\tau/2mk_\text{F} \sim -10^{-3}$. This means that the result is not
far from the converged solution after the first iteration. To maintain
the Thouless criterion, we subtract numerically this $\tau$ from the
inverse $T$ matrix, when calculating the pair dispersion via
$\chi^{\text{R}}({\mathbf{q}},\Omega)$.

Note that Eqs.~(\ref{eq:susconv}), (\ref{eq:tpgconv}),
(\ref{eq:Sigmapgconv}), and (\ref{eq:Gfuncconv}) form a
self-consistent loop, as indicated by the line linking
$\text{Im}\,G^{\text{R}}({\mathbf{k}},\omega)$ and
$\chi^{\text{R}}({\mathbf{q}},\Omega)$ in Fig.~\ref{fig:flowchart}.
Combined with the number equation,
\begin{equation}
  n =\sumk \int \frac{\d\omega}{\pi} A(\mathbf{k},\omega) f(\omega)\,,
  \label{eq:num2}
\end{equation}
and the Thouless criterion, these equations can be solved iteratively
for $(\mu, T_{\text{c}})$ until convergence is reached, yielding
$\bar{E}_{\text{Hartree}}$, $A({\mathbf{k}},\omega)$, and
$B({\mathbf{q}},\Omega)$ for each iteration.  However, the multifold
integrations in Eqs.~(\ref{eq:susconv}) and (\ref{eq:Sigmapgconv}),
especially with sharp peaks below $T_\text{c}$, are very demanding in
computational resources, so that fully self-consistent calculations
are deferred to a future work, which may leverage advanced numerical
techniques to address these challenges. Nonetheless, the smallness of
$\tau$ after the first iteration suggests that the resulting
spectral function, which  captures the lifetime effects of both
fermions and pairs, can already be used for comparison with the recent
microwave spectroscopic measurements \cite{Li2024N}. It is worth
mentioning that in this iterative numerical approach, the parameter
$\Delta_\text{pg}$, which is an important feature of the pseudogap
approximation in Eq.~\eqref{eq:Sigmapgappro}, no longer appears, while
it can still be extracted from the resulting spectral function or DOS.

\section{NUMERICAL RESULTS AND DISCUSSIONS}
\label{sec:3}

In this section, we  present our representative results on key
physical properties of a homogeneous Fermi gas across the BCS-BEC
crossover, including the average Hartree self-energy, the physical
chemical potential, the fermion spectral function, and the DOS, using
the iterative framework described above.  We  extract the pairing
gap from EDCs and compare with experimental
data.  Furthermore, we  also show the behavior of the pair
spectral function $B(\mathbf{q}, \Omega)$.

\subsection{Hartree Self-Energy and the Chemical Potential}

\begin{figure}
\centering
\includegraphics[clip,width=3.4in]{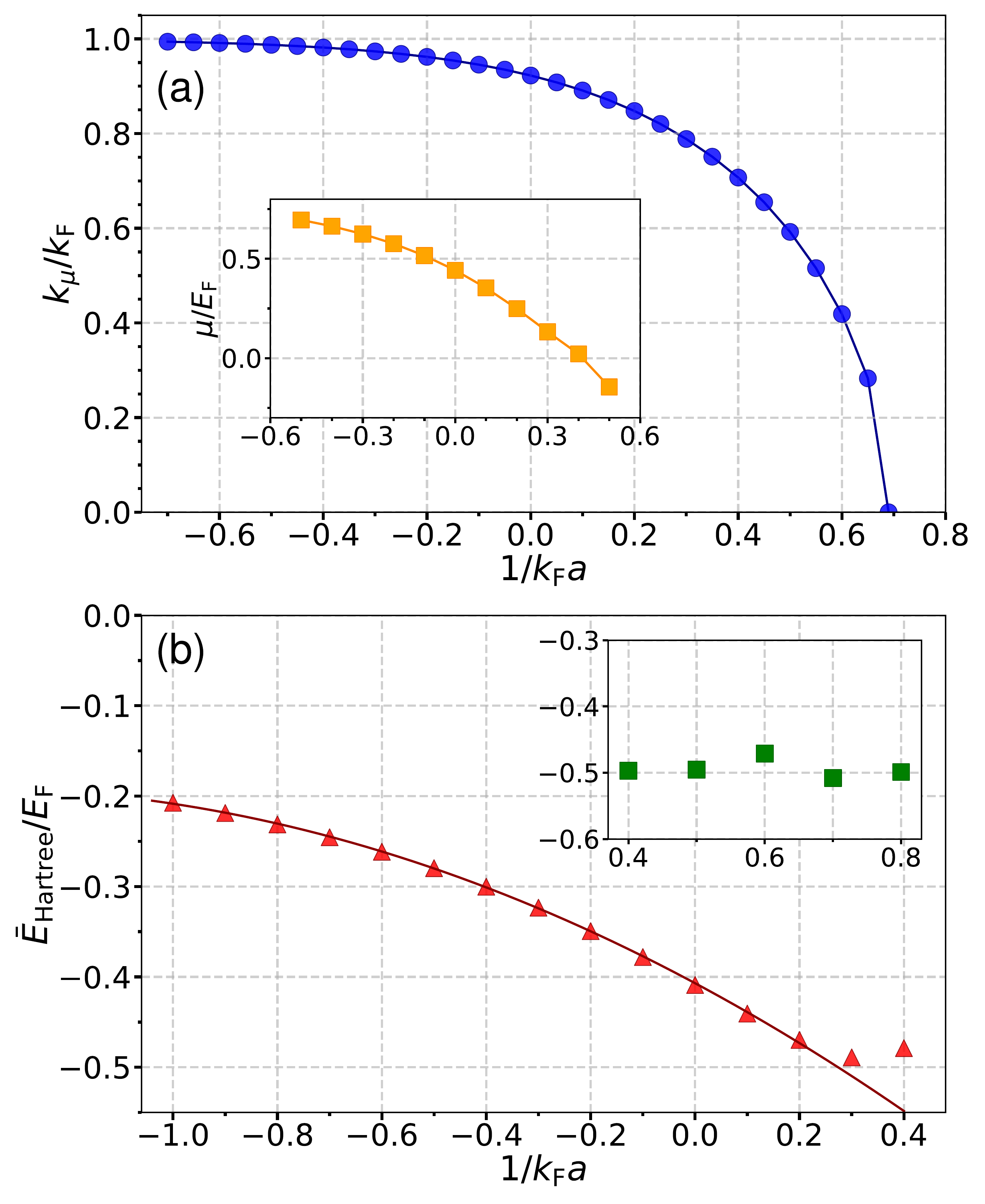}
\caption{ (a) Wave vector $k_\mu$ versus $1/k_{\text{F}}a$ at
  $T_{\text{c}}$ in the fermionic regime where a Fermi surface is
  present.  Shown in the inset is the physical chemical potential
  $\mu(T_{\text{c}})$ in this regime.  (b) Average Hartree energy
  $\bar{E}_{\text{Hartree}}$ as a function of $1/k_{\text{F}}a$ at
  $T_{\text{c}}$.  The data points fit nicely to a second-order polynomial,
  indicating a smooth evolution from the BCS to the BEC regime.  The
  inset shows $\bar{E}_{\text{Hartree}}$ extracted from the hole
  branch dispersion in the near-BEC regime.  }
\label{fig:hartreekl}
\end{figure} 

Shown in Fig.~\ref{fig:hartreekl}(a) is the wave vector $k_\mu$ on the
Fermi surface (where the BCS-like quasiparticle dispersions exhibit
back-bending) at $T_{\text{c}}$, as a function of the interaction
strength $1/k_{\text{F}}a$.  In the BCS regime, $k_\mu$ decreases
gradually from the Fermi momentum $k_{\text{F}}$ in the noninteracting
limit as $1/k_{\text{F}}a$ increases.  Upon entering the unitary
regime, $k_\mu$ drops rapidly to zero near $1/k_{\text{F}}a \approx
0.7$, signifying the disappearance of the Fermi surface.  Plotted in
the inset is the physical chemical potential $\mu(T_{\text{c}})$,
which passes zero at $1/k_{\text{F}}a \approx 0.4$.  Shown in
Fig.~\ref{fig:hartreekl}(b) is the average Hartree energy
$\bar{E}_{\text{Hartree}}$ versus $1/k_\text{F}a$ at $T_{\text{c}}$,
computed via
$\bar{E}_{\text{Hartree}}=\text{Re}\Sigma^\text{R}_\text{pg}(k_\mu,0)$.
We find that the data points of $\bar{E}_{\text{Hartree}}$ follow
nicely a second-order polynomial fit, manifesting a smooth evolution
from the BCS to the BEC regime.  In the BEC regime,
$\bar{E}_{\text{Hartree}}$ remains nearly constant at
$-0.5E_{\text{F}}$, as shown in the inset, which is extracted by
subtracting the off-diagonal self-energy contributions using the
hole-branch dispersion.  Note that here the Hartree energy includes
contributions beyond the leading-order term $ng$. In the zero-range
contact potential limit, $g$ is renormalized down to
$0^-$. Nonetheless, in agreement with the Galitskii expansion, we find
it proportional to $k_\text{F}a$ in the BCS limit, and it varies
roughly linearly as a function of $ 1/k_\text{F}a$ in the unitary
regime.  At $1/k_{\text{F}}a\approx 0.7$, where $\mu'=0$, we find that
the physical $\mu = \bar{E}_{\text{Hartree}} \approx
-0.5E_{\text{F}}$, consistent with the results of the 
Luttinger-Ward formalism \cite{Haussmann2009PRA} and the $\epsilon$
expansion \cite{Nishida2007PRA}.

\begin{figure}
\centering
\includegraphics[clip,width=3.2in] {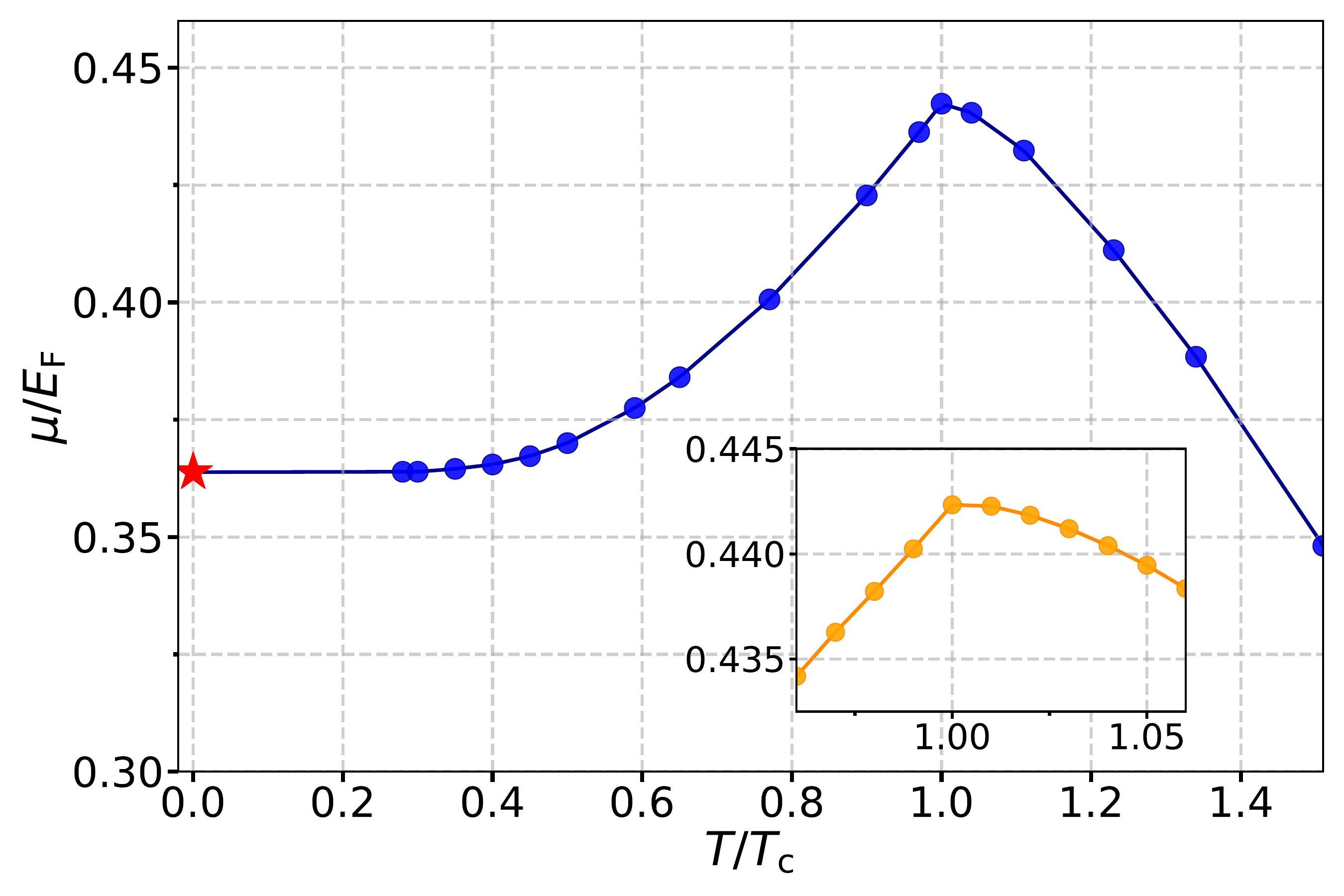}
\caption{ Physical chemical potential $\mu$ of a unitary Fermi gas as
  a function of temperatures both above and below $T_c$.  The red star
  marks the extrapolated Bertsch parameter $\xi=0.364$ at $T=0$.  The
  inset zooms in near $T_\text{c}$, highlighting a rather abrupt change in
  the slope of $\mu(T)$ across $T_c$.}
\label{fig:mu}
\end{figure}

Next, we present in Fig.~\ref{fig:mu} the physical chemical potential
$\mu$ at unitarity as a function of $T/T_{\text{c}}$.  It reaches a
maximum at $T_{\text{c}}$, then gradually decreases as
$T/T_{\text{c}}$ falls, due to the opening of the pairing gap, and
approaches a zero $T$ asymptote for $T/T_{\text{c}} < 0.3$. This
nonmonotonic $T$ dependence is typical of a mean-field BCS
superconductor.  An abrupt change in the slope of $\mu$ at
$T_{\text{c}}$ aligns with previous thermodynamic measurements
\cite{Ku2012S}.  At unitarity, its ground-state value is characterized
by the Bertsch parameter $\xi = \mu/E_\text{F}$ at $T = 0$.
Extrapolating $\mu$ to zero temperature yields $\xi=0.364$, consistent
with results from experiment \cite{Ku2012S,Zurn2013PRL}, Monte Carlo
calculations \cite{Carlson2011PRA,Endres2013PRA,Pessoa2015PRA} and
$\epsilon$ expansion \cite{Nishida2009PRA}. It is interesting to note
that, despite the opening of a pseudogap already above $T_c$ at
unitarity, the maximum of $\mu$ is still observed roughly at $T_c$
rather than at a higher temperature. This manifests the different
effects on $\mu$ between a true superconducting gap and a pseudogap.

\subsection{Fermion Spectral Function}

\begin{figure}
\centering
\includegraphics[clip,width=3.4in]{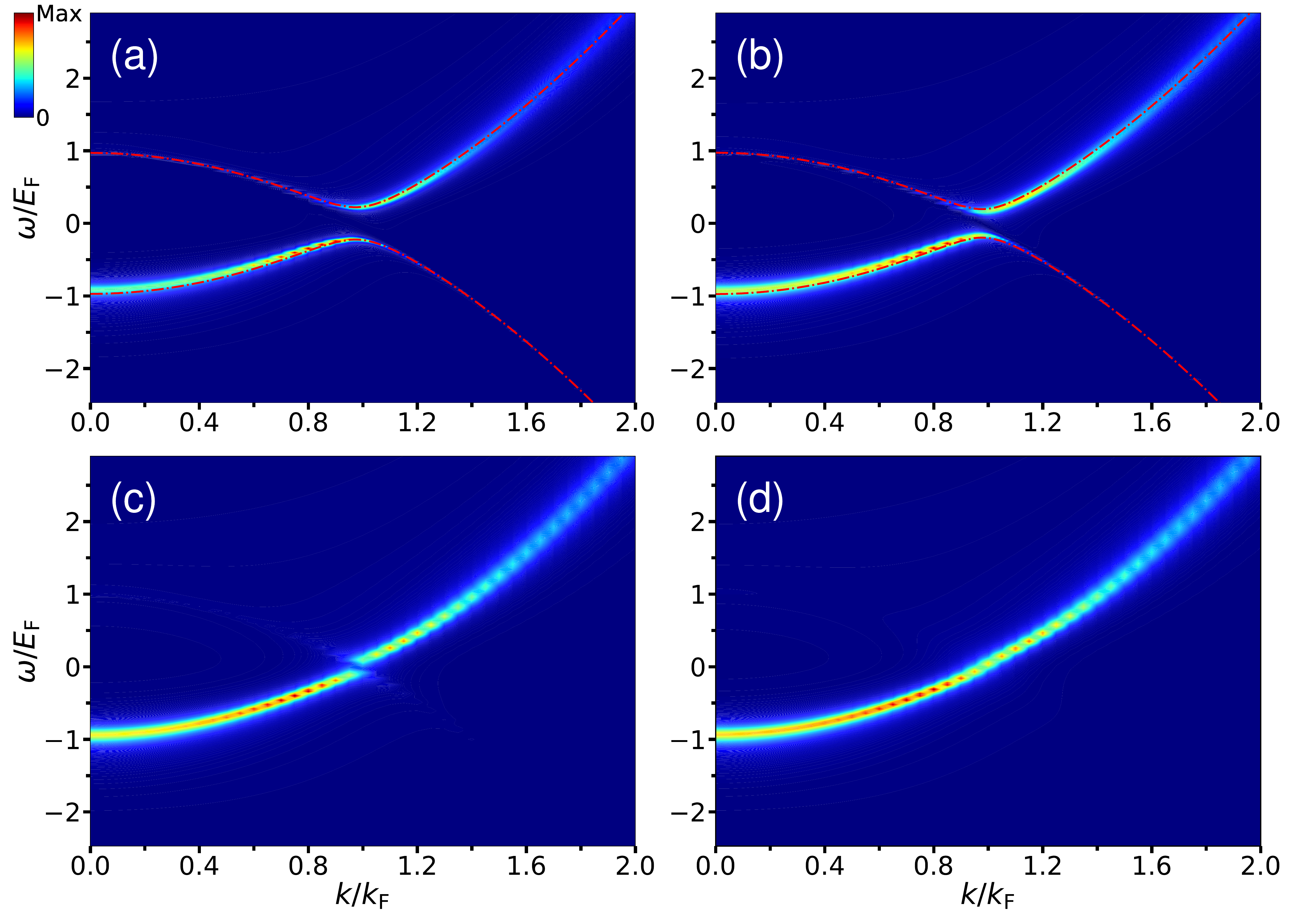}
\caption{ Contour plots of the spectral function
  $A(\mathbf{k},\omega)$ at (a) $T/T_{\text{c}} = 0.7$, (b) $0.9$, (c)
  $1$, and (d) $1.1$ for $1/k_{\text{F}}a = -0.4$.  Red dot-dashed
  lines in (a) and (b) show the dispersion curves from the initial
  self-consistent solutions under the pseudogap approximation.  }
\label{fig:AkwBCS}
\end{figure}

Shown in Fig.~\ref{fig:AkwBCS} are the contour plots of the spectral
function $A(\mathbf{k},\omega)$ as a function of $k = |\mathbf{k}|$
and $\omega$ in the BCS regime with $1/k_{\text{F}}a = -0.4$ for
$T/T_{\text{c}}$ ranging from $0.7$ to $1.1$. At lower temperatures
below $T_c$, as shown in (a) and (b), the sharp double peaks in the spectral
intensity  for fixed $k$ near the Fermi level, $k \approx k_\mu$,
indicate stable Cooper pairing around the Fermi surface in the
superfluid phase of the weak-coupling regime. Two branches both
exhibit a back-bending behavior, manifesting clear-cut particle and
hole branches of BCS-like dispersions caused by Cooper pairing.  For
comparison, we overlay on top of the intensity map the dispersion
curves (red dot-dashed lines) derived from the self-consistent
solutions of Eqs.~\eqref{eq:num}, \eqref{eq:pg}, and \eqref{eq:gapph}
under the pseudogap approximation, which align well with the spectral
peaks, validating the approximation in Eq.~\eqref{eq:Sigmapgappro}. As
$T$ increases to $T_{\text{c}}$ in Fig.~\ref{fig:AkwBCS}(c), the
pairing gap shrinks, as evidenced by the particle and hole branches
moving toward each other. At $T/T_{\text{c}} = 1.1$ above $T_c$ in
Fig.~\ref{fig:AkwBCS}(d), the spectral intensity peak near $k_\mu$ is
only slightly suppressed, and a nearly quadratic dispersion emerges,
resembling that of a free fermion. This indicates that the pseudogap
effect is rather weak at this near-BCS interaction strength.

\begin{figure}
\centering
\includegraphics[clip,width=3.4in]{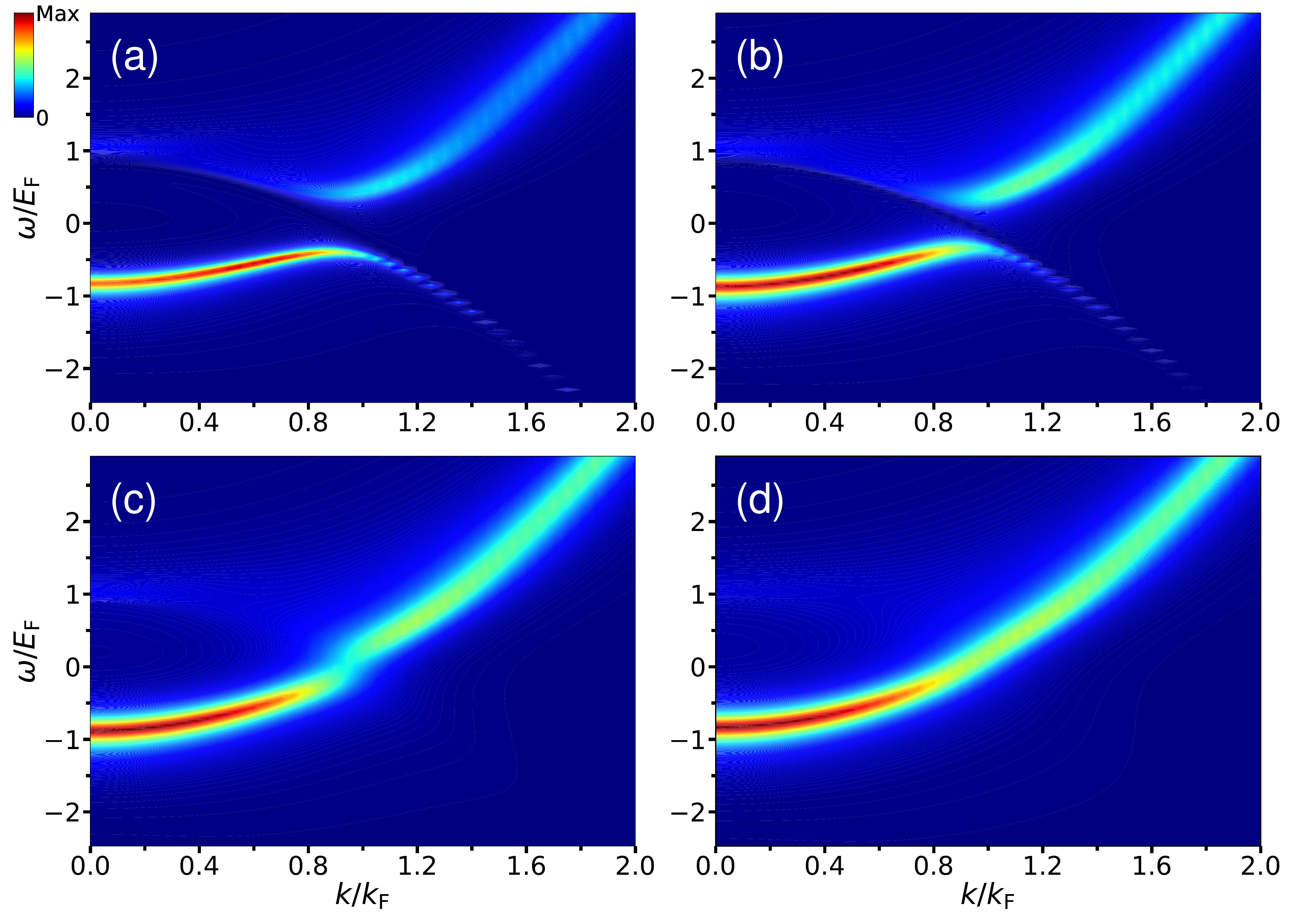}
\caption{
  Contour plots of the spectral function $A(\mathbf{k},\omega)$ at (a) $T/T_{\text{c}} = 0.7$, (b) $0.9$, (c) $1.1$, and (d) $1.3$ for $1/k_{\text{F}}a = 0$.
}
\label{fig:Akwunitary}
\end{figure}

Next, we present in Fig.~\ref{fig:Akwunitary} contour plots of
$A(\mathbf{k},\omega)$ in the $(k = |\mathbf{k}|, \omega)$ plane at
unitarity around $T_{\text{c}}$. Below $T_{\text{c}}$, in panels (a)
and (b), we observe two BCS-like dispersions with back-bending,
similar to the BCS case. Above $T_{\text{c}}$, in
Fig.~\ref{fig:Akwunitary}(c), these dispersions hybridize into an
S-shaped curve. The upper branch at low momenta is clearly visible,
and a significant pseudogap can be identified at the back-bending
point (i.e., the Fermi level) at this temperature. The dispersions
exhibit significant broadening, driven by the higher absolute
temperature at unitarity due to a larger $T_{\text{c}}$. At higher
$T/T_{\text{c}} = 1.3$ above $T_c$ in Fig.~\ref{fig:Akwunitary}(d), a
subtle S-shaped dispersion in $A(\mathbf{k},\omega)$ reveals a
persistent pseudogap arising from residual pairing; the dispersion
deviates from a simple parabola, along with weak but visible
intensities of the spectral weight of the upper-branch dispersion.

\begin{figure}
\centering \includegraphics[clip,width=3.4in]{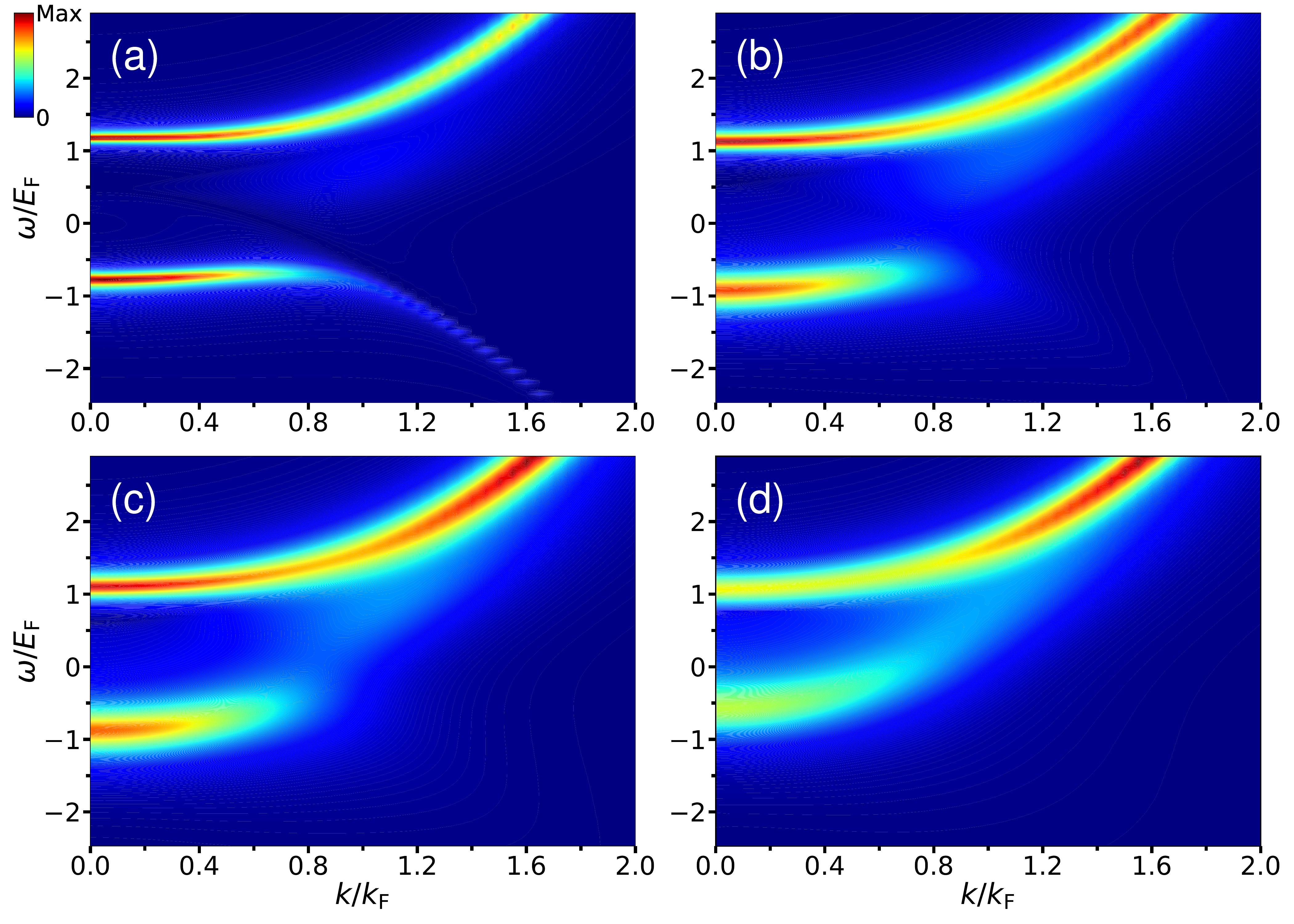}
\caption{ Contour plots of the spectral function
  $A(\mathbf{k},\omega)$ at (a) $T/T_{\text{c}} = 0.7$, (b) $1.1$, (c)
  $1.3$, and (d) $1.8$ for $1/k_{\text{F}}a = 0.4$ on the BEC side of
  the unitarity.  }
\label{fig:AkwBEC}
\end{figure}

Shown in Fig.~\ref{fig:AkwBEC} is $A({\mathbf{k}},\omega)$ as a
function of $k=|{\mathbf{k}}|$ and $\omega$ with $1/k_{\text{F}}a=0.4$
on the BEC side of unitarity, at $T/T_{\text{c}}=0.7$, $1.1$, $1.3$,
and $1.8$ for panels (a)–(d), respectively, where a large pairing gap
arises due to strong interactions.  The upper branch exhibits higher
spectral weight than in the BCS and unitary regimes, reflecting
stronger particle-hole mixing due to a larger gap.  At
$T/T_{\text{c}}=0.7$ in panel (a), the particle branch reaches the
largest spectral weight at small $k$, with a broad scattering
continuum background, lacking a clear back-bending point, thus
challenging the description in terms of a BCS-like dispersion.  The
hole branch, however, exhibits a back-bending, with its maximum weight
at $k = 0$.  As the temperature increases from panel (b) to (d), the
two branches come closer and have a tendency to merge. The
spectral weight of the particle branch at small $k$ remains large due
to the presence of a large pseudogap even at $T/T_{\text{c}}=1.8$.  At
even higher temperatures (not shown), the scattering continuum
background continues to grow, and the gap decreases further so that
the spectral weight at small $k$ of the particle branch is gradually
shifted to the hole branch, and eventually one is left with a nearly
quadratic free-fermion dispersion. Similar high $T$ behavior is also
observed in rf spectral calculations based on the $G_{0}G_{0}$
\cite{Reichl2015PRA} and $GG$ schemes
\cite{Haussmann2009PRA,Johansen2024PRA,Dizer} of $T$-matrix
approximations.  The observation that the back-bending point in the
hole branch shifts to lower $k$ than in the near-BCS and unitary cases
reflects a reduced $\mu'$ or shrunken Fermi surface at this stronger
interaction strength.  The asymmetry between the particle and hole
branches of the dispersions reflects that the actual quasiparticle
dispersions in the presence of strong pairing fluctuations are more
complicated than an oversimplified BCS form.

\subsection{Single-Particle Density of States}

\begin{figure}
\centering
\includegraphics[clip,width=3.2in]{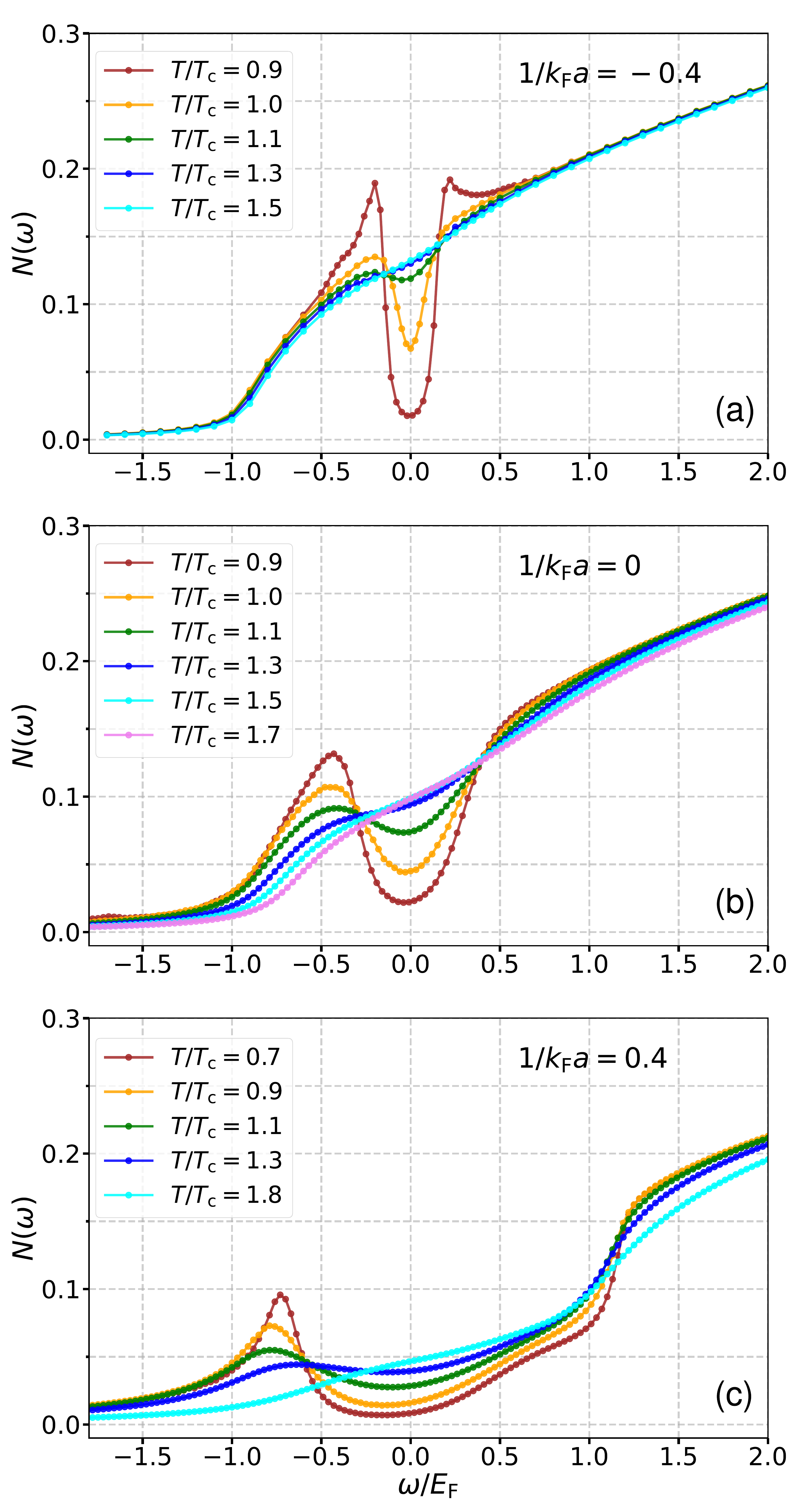}
\caption{Temperature evolution of the DOS $N(\omega)$ with a series of
  increasing $T/T_{\text{c}}$ from below to above $T_c$ for (a)
  $1/k_{\text{F}}a = -0.4$, (b) $0$, and (c) $0.4$ from weak to
  strong interactions.  }
\label{fig:DOS}
\end{figure}

The DOS $N(\omega)$, derived from the spectral function
$A({\mathbf{k}},\omega)$, is given by
\begin{equation}
\label{eq:DOS}
N(\omega) = \sum_{\mathbf{k}} A(\mathbf{k},\omega)\,.
\end{equation}
In Fig.~\ref{fig:DOS}, we show the temperature evolution of
$N(\omega)$ as a function of $\omega$ from below to above
$T_{\text{c}}$ at (a) $1/k_{\text{F}}a = -0.4$, (b) $0$, and (c)
$0.4$, representing the near-BCS, unitary, and near-BEC cases,
respectively.  Below $T_{\text{c}}$, a significant depletion near
$\omega=0$ (on the Fermi surface) results from a pairing gap.  This
DOS depletion persists above $T_{\text{c}}$ in all cases. For
$1/k_{\text{F}}a = -0.4$, the DOS roughly returns to normal without a
pseudogap at $T/T_c = 1.5$, even though one might argue that a weak
broad depression is still discernible. At unitarity, such depression
persists up to $1.7T_c$.  In the near-BEC regime shown in
Fig.~\ref{fig:DOS}(c), a pseudogap persists even above $T/T_{\text{c}}
\approx 1.8$ (see Fig.~\ref{fig:AkwBEC}).  The energy width of the
depletion at $T_{\text{c}}$ expands with increasing $1/k_{\text{F}}a$,
reflecting a stronger pseudogap arising from pairing fluctuations.  As
$T$ increases, the DOS is gradually filled in, resembling that
observed in high-$T_{\text{c}}$ superconductors \cite{Timusk1999RPP}.
Above $T_{\text{c}}$, finite-momentum pairs become short-lived and
break apart at high $T$, so that the pseudogap decreases and the DOS
becomes filled in until it looks like its non-interacting counterpart
around the pair formation temperature $T^{*}$.  Note that the
interaction-induced spectral broadening will give rise to nonzero DOS
even below the free-fermion band bottom $\omega = -\mu$.

\subsection{Comparison with Recent Experiments}

\begin{figure}
\centering
\includegraphics[clip,width=3.4in]{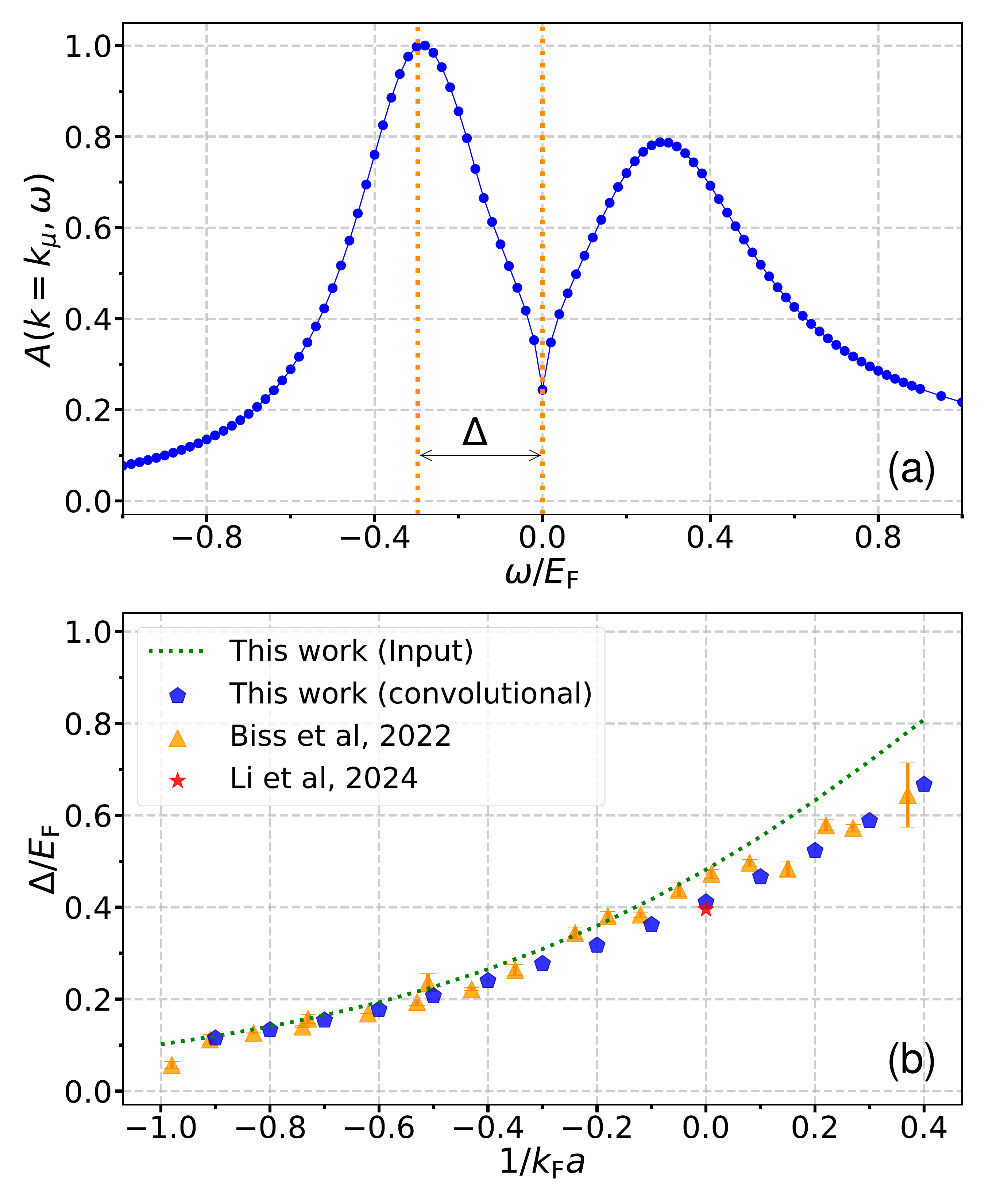}
\caption{ (a) Normalized EDC of $A(\mathbf{k},\omega)$ at $k_\mu$ for
  a unitary Fermi gas at $T_{\text{c}}$. The pairing gap $\Delta$ can
  be directly obtained from the separation between the peak and the
  central minimum in the hole branch. (b) Comparison between the gap
  $\Delta$ extracted from numerically generated EDCs (blue pentagons,
  labeled ``convolutional'', at $T/T_{\text{c}} = 0.5$) and from
  experimental data, as a function of $1/k_{\text{F}}a$. The green
  dashed line shows the initial input from self-consistent solutions
  under the pseudogap approximation, labeled ``input''. Orange
  triangles and the red star represent experimental results from
  Refs.~\cite{Biss2022PRL} (temperature not specified)
  and \cite{Li2024N} (at $T/T_{\text{c}} = 0.77$), respectively.  }
\label{fig:delta}
\end{figure}

In Fig.~\ref{fig:delta}, we compare the pairing gap $\Delta$ extracted
from our computed spectral functions with experimental data.  Plotted
in Fig.~\ref{fig:delta}(a) is the normalized EDC of $A(\mathbf{k},
\omega)$ at $k_\mu$ of a unitary Fermi gas at $T_{\text{c}}$ as a
function of $\omega$.  The EDC's two peaks reflect quasiparticle
energies. We extracted $\Delta$ from the separation between the peak
and the central minimum of the hole branch, as the particle branch
deviates from the BCS-like dispersion in the strong coupling regime as
depicted in Fig.~\ref{fig:AkwBEC}. In Fig.~\ref{fig:delta}(b), we
compare the excitation gap extracted from our numerical data at
$T/T_{\text{c}}=0.5$ and from experiments as a function of
$1/k_{\text{F}}a$.  The green dashed line represents the
self-consistent solution from Eqs.~(\ref{eq:num}), (\ref{eq:pg}) and
(\ref{eq:gapph}), and the blue pentagons denote $\Delta$ extracted
from our computed EDCs.  Orange triangles and the red star represent
experimental results from Bragg spectroscopy \cite{Biss2022PRL} and
microwave spectroscopy ($T/T_{\text{c}}=0.77$) \cite{Li2024N},
respectively.  In the weak coupling regime, the gap values extracted
from our numerical EDCs (blue pentagons) are very close to the initial
self-consistent solution under the pseudogap approximation (green
dashed line), consistent with Fig.~\ref{fig:AkwBCS}. This is because
finite-momentum pair contributions are weak so that the fermion self
energy is dominated by the mean-field BCS order parameter.  As the
interaction becomes stronger, the EDC-extracted gap starts to deviate
significantly from the solution under the pseudogap approximation;
here contributions of finite-momentum pairs become important, so that
the pseudogap approximation in Eq.~(\ref{eq:Sigmapgappro}) becomes
quantitatively less accurate.  The EDC-extracted gaps agree well with
the experimental results from the excitation spectrum of an ultracold
$^6\text{Li}$ gas \cite{Biss2022PRL} at different interaction
strengths and the value at unitarity from Ref.~\cite{Li2024N} (red
star, measured at a higher temperature $T/T_{\text{c}}=0.77$).

\begin{figure*}
\centering
\includegraphics[clip,width=7in]{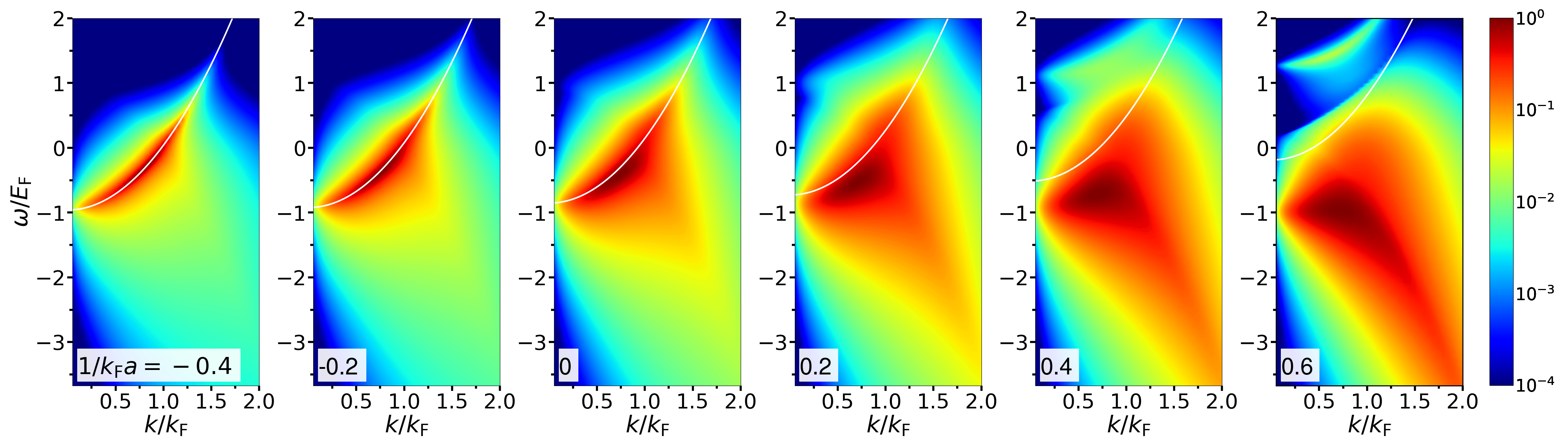}
\caption{Intensity maps of the numerically simulated
  $\mathbf{k}^{2}A(\mathbf{k},\omega)f(\omega)$ of a homogeneous Fermi
  gas slightly above $T_c$ at $T/T_{\text{c}} = 1.1$ for a series of $1/k_{\text{F}}a$
  across the BCS-BEC crossover. The white lines represent the
  Hartree-shifted free fermion dispersion $\xi_{\mathbf{k}} =
  \mathbf{k}^{2}/2m - \mu'$. Note that the color-coding for the
  spectral intensity is in a logarithmic scale. }
\label{fig:Akwofdiffka}
\end{figure*}

Previous momentum-resolved rf spectra of Fermi gases at the 3D trap
center revealed a breakdown of the Fermi liquid description at strong
couplings~\cite{Sagi2015PRL}.  In Fig.~\ref{fig:Akwofdiffka}, we
present contour plots of our computed angle-integrated rf spectral
intensity
$I(\mathbf{k},\omega)=\mathbf{k}^{2}A({\mathbf{k}},\omega)f(\omega)$
as a function of $k=|{\mathbf{k}}|$ and $\omega$, at
$T/T_{\text{c}}=1.1$, which is the normal state slightly above $T_c$,
for a range of $1/k_{\text{F}}a$ from $-0.4$ to $0.6$, from weak to
strong couplings.  On a logarithmic scale, the spectra exhibit a
widespread incoherent spectral intensity distribution at negative
$\omega$, which increases with interaction strength. This can be
easily understood. First, the larger gap at a stronger interaction
leads to stronger particle-hole mixing and a wider spread of the
spectral weight into higher momenta above $k_F$, i.e., a wider spread
of $\vk^2$ for the hole branch. At the same time, a stronger
interaction causes larger spectral broadening and hence a much larger
spectral spread as a function of frequency. We do not see the enhanced
intensity near the Hartree-shifted ``free'' fermion dispersion
$\xi_{\mathbf{k}}=\mathbf{k}^{2}/2m-\mu'$ (white line) that was
observed experimentally~\cite{Sagi2015PRL}, which was likely caused by
the trap inhomogeneity such that a free fermion signal arose at the
trap edge. This signal was overlapped on top of the particle branch of
the signals from the rest of the trap. Indeed, the focused rf
beam used in Ref.~\cite{Sagi2015PRL} necessarily passed through the
trap edge and was expected to have led to the free fermion signal. In
addition, possible non-equilibrium effects might also have contributed
to the free-fermion signal. Indeed, such a free fermion signal was not
observed at unitarity in the experiment of Li et
al.~\cite{Li2024N}. Note that the data at unitarity shown here is the
same set as in Fig.~2(c) of Ref.~\cite{letter}, when divided by
$f(\omega)$. The logarithmic scale also makes the intensity map appear
rather different.

It is useful to also compare our theoretically extracted pseudogap with
competing $T$-matrix based calculations. Using the $G_{0}G_{0}$
scheme, Refs.~\cite{Strinati2012} and \cite{Reichl2015PRA} found a
pseudogap of about $0.85E_F$ at $T_c$ and $0.73E_F$ at $1.1T_c$, both
of which seemed to be overly large compared to experiment. In contrast,
the gap seems to vanish at the Fermi level in the $GG$ scheme of the
$T$-matrix
approximation~\cite{Tchernyshyov1997,Haussmann2009PRA,Johansen2024,Enss2024}.

\subsection{Pair Spectral Function}

\begin{figure}
\centering
\includegraphics[clip,width=3.4in]{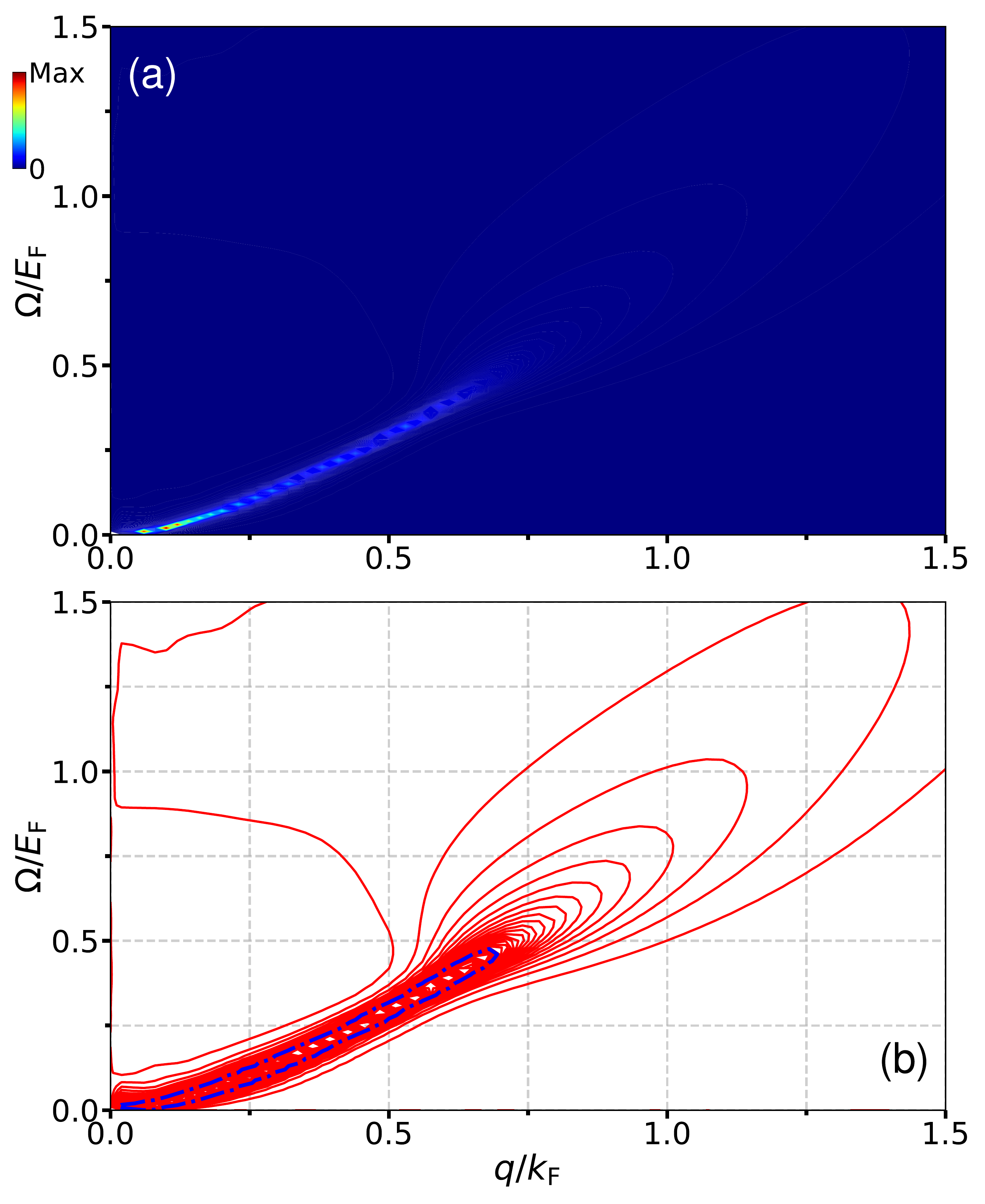}
\caption{ (a) Intensity map and (b) contour lines of the pair spectral
  function $B(\mathbf{q},\Omega)$ at unitarity for $T/T_{\text{c}} =
  0.5$ in the superfluid phase.  The blue dot-dashed line in (b)
  denotes the saturation threshold.  }
\label{fig:Bqw0}
\end{figure}

Finally, in this subsection, we investigate the behavior of the pair
spectral function $B({\mathbf{q}},\Omega)$ and analyze the excitation
spectrum of finite-momentum pairs.  As shown in Fig.~\ref{fig:Bqw0}(a)
for the unitary case at $T/T_{\text{c}}=0.5$, the pair spectral
intensity map of $B({\mathbf{q}},\Omega)$ in the
$(q=|{\mathbf{q}}|,\Omega)$ plane reveals a well-defined parabolic
dispersion at low $\Omega$, indicative of long-lived pair states.
The contour lines in Fig.~\ref{fig:Bqw0}(b) show that the dispersion
 broadens rapidly for $\Omega > 0.7E_\text{F}$, so that
finite-momentum pairs become short-lived and diffusive at these high
energies.  The blue dot-dashed line marks the saturation threshold,
above which spectral weight is truncated in the color-coding in
subsequent figures.

\begin{figure}
\centering
\includegraphics[clip,width=3.4in]{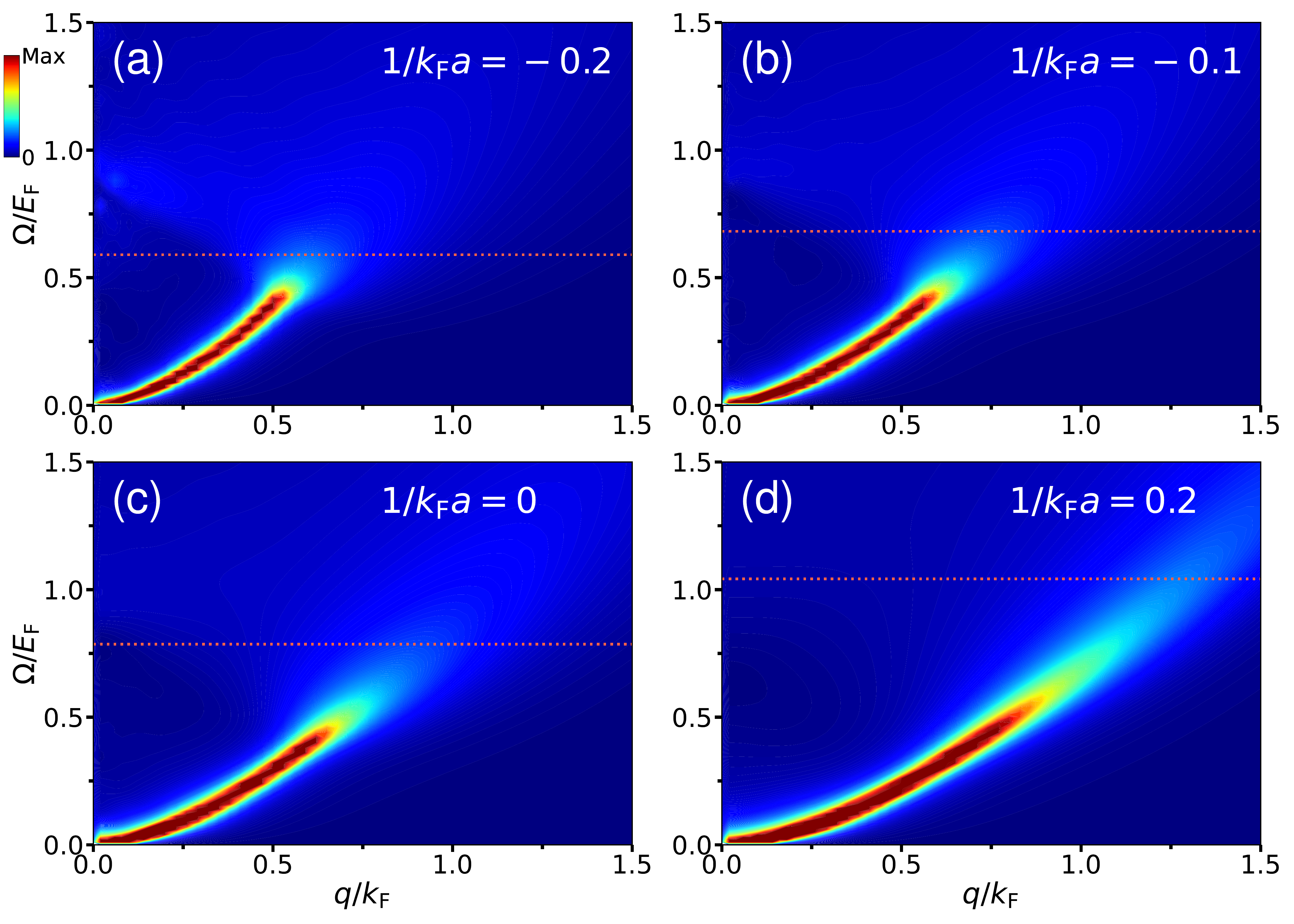}
\caption{Evolution of the spectral intensity map of
  $B(\mathbf{q},\Omega)$ computed at $T/T_{\text{c}} = 0.7$ with the
  interaction strength from weak to strong: (a) $1/k_{\text{F}}a =
  -0.2$, (b) $-0.1$, (c) $0$, and (d) $0.2$.  The red dashed lines
  mark the energy $\Omega = 2\Delta$.  }
\label{fig:Bqw1}
\end{figure}

In Fig.~\ref{fig:Bqw1}, we present the behaviors of the spectral
intensity map of $B(\mathbf{q},\Omega)$ in the $(q, \Omega)$ plane for
different interaction strengths $1/k_{\text{F}}a = -0.2$, $-0.1$, $0$,
and $0.2$, from weak to strong. The data were computed at
$T/T_{\text{c}} = 0.7$ for the superfluid phase, and the spectral
weight is truncated at the saturation threshold in the color coding.
The parabolic pair dispersion at low $\Omega$ becomes softer with
increasing interaction strength, reflecting an increasing effective
pair mass. Above $\Omega = 2\Delta$ (red dashed lines, where $\Delta$
is taken from Fig.~\ref{fig:delta}), the spectral weight rapidly
spreads out in both momentum and frequency. The dispersion becomes
diffusive and no longer well-defined for $\Omega > 2\Delta$. With
increasing interaction strength, long-lived pairs can exist in a
progressively larger range of momentum and frequency, due to the
increasing pairing gap. This suggests that $2\Delta$ may serve
roughly as the pair-breaking energy.

\begin{figure}
\centering
\includegraphics[clip,width=3.4in]{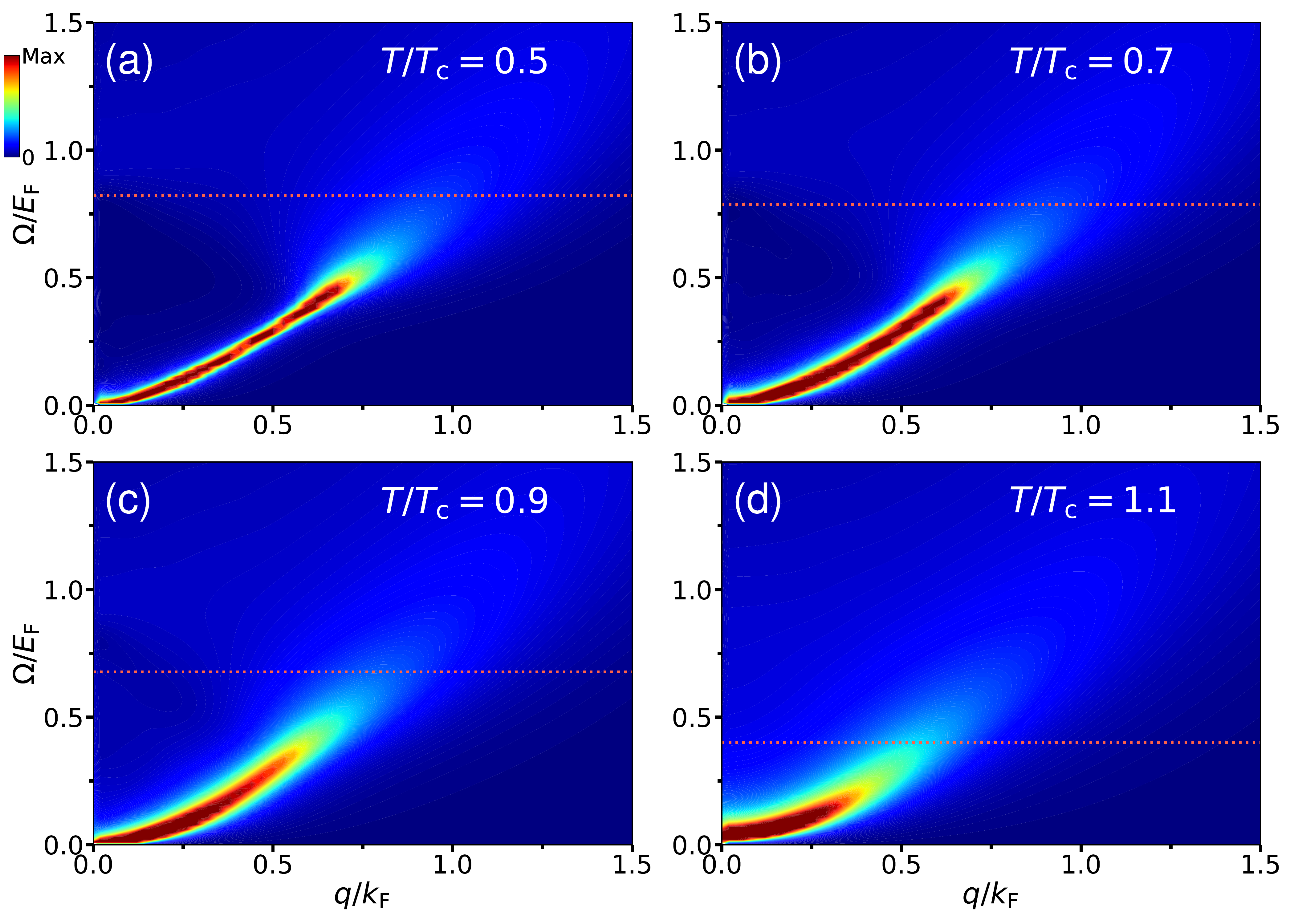}
\caption{ Temperature evolution of $B(\mathbf{q},\Omega)$ at unitarity
  for (a) $T/T_{\text{c}} = 0.5$, (b) $0.7$, (c) $0.9$, and (d)
  $1.1$. Red dashed lines mark $\Omega = 2\Delta$, and the spectral
  intensity is truncated at the saturation threshold given in
  Fig.~\ref{fig:Bqw0}. }
\label{fig:Bqw2}
\end{figure}

Finally, we show in Fig.~\ref{fig:Bqw2} the temperature evolution of
the pair spectral function $B(\mathbf{q},\Omega)$, calculated at
unitarity for $T/T_{\text{c}} = 0.5$, $0.7$, $0.9$, and $1.1$.  Below
$T_{\text{c}}$, a well-defined gapless parabolic dispersion is
present, reflecting the long lifetime of the pairs, and thus a sharp
peak at low $\Omega$ for fixed small $q$. As $T/T_{\text{c}}$
increases, the spectral peak becomes broader, the pair lifetime becomes
shorter, and finite-momentum pairs, driven by thermal excitations,
contribute to a larger pseudogap \cite{Chen2014FP}. Above
$T_{\text{c}}$ in Fig.~\ref{fig:Bqw2}(d), the dispersion has a gap,
albeit small, at $q=0$, given by the absolute value of the negative
pair chemical potential $\mu_{\text{p}}$. As the temperature
increases, the pair-breaking energy scale $2\Delta$ decreases.  It is
expected that as $T/T_{\text{c}}$ increases further, the spectral peak
will become so broad that the pairs will no longer be well-defined,
with only a diffusive dispersion. In this case, one no longer has a
pseudogap.  Recent Keldysh-based studies
\cite{Johansen2024PRA,Enss2024PRA} also reported broadened pair
spectra in the normal state but missed pair-breaking processes due to
the short pair lifetime at high temperature.

\section{CONCLUSIONS}
\label{sec:4}

To summarize, we have studied the spectra of ultracold Fermi gases
across the BCS-BEC crossover using an iterative framework that calculates
the fermion and pair spectral functions. Going beyond
the previous pseudogap approximation, we employed a full numerical
convolution to compute the self-energy accurately. From this
framework, we determined the average Hartree energy and the physical
chemical potential, extracted the Bertsch parameter, and obtained
reliable spectral functions and density of states.

The computed fermion spectral intensity maps reveal broadened
quasiparticle dispersions, indicating a  pseudogap that
grows with interaction strength from the BCS to the BEC regime --- a
finding confirmed by the DOS. Notably, the pseudogap is more
pronounced in the DOS than in the spectral function itself.  The pair
spectral intensity maps exhibit a well‑defined finite‑momentum pair
dispersion at low energies; above the pair‑breaking scale (roughly
$2\Delta$), the dispersion becomes diffusive. Thermal excitations of
these finite‑momentum pairs contribute to the pseudogap. The $2\Delta$
scale was also extracted from temperature‑dependent quasiparticle
lifetime analysis, as detailed in the companion paper
\cite{letter}.

Crucially, the pairing gap extracted from our computed EDCs and the
simulated rf spectral intensity shows quantitative agreement with
recent momentum‑resolved microwave spectroscopy measurements on a
homogeneous unitary Fermi gas \cite{Li2024N} (see also the companion
paper \cite{letter}). It also agrees with recent Bragg spectroscopy
data \cite{Biss2022PRL} and momentum-resolved rf spectroscopy data in
a trap \cite{Sagi2015PRL}. These agreements provide strong support for
the pairing origin of the pseudogap in strongly interacting Fermi
gases and demonstrate that strong pairing can generate a pseudogap --- a
picture that is arguably applicable to high‑$T_c$ superconductors as
well.

Although our iterative procedure has not yet been carried to full
self‑consistency for $(\mu, \Delta)$, it correctly captures the
essential effects of spectral broadening and the Hartree self-energy,
thereby offering a detailed microscopic understanding of the pseudogap
and the spectral behavior observed in ultracold Fermi gases. In
principle, the pair momentum distribution can be measured via a rapid
magnetic field sweep that converts pairs into tightly bound molecules,
followed by time‑of‑flight expansion and optical imaging.

\section{ACKNOWLEDGMENTS}
\label{sec:5}
This work was supported by the Quantum Science and Technology - National
Science and Technology Major Project (Grant No. 2021ZD0301904) of
China.

\bibliographystyle{apsrev4-2}
\bibliography{Refs.bib}

@article{Leggett2006NP,
  Author = {Leggett, AJ},
  Title = {What DO we know about high ${T}_{c}$?},
  Journal = {Nature Physics},
  Year = {2006},
  Volume = {2},
  Number = {3},
  Pages = {134-136},
  Month = {MAR},
  DOI = {10.1038/nphys254},
  ISSN = {1745-2473},
  Unique-ID = {WOS:000236128200002},
}

@article{Nishida2009PRA,
  title = {Ground-state energy of the unitary Fermi gas from the $\epsilon$ expansion},
  author = {Nishida, Yusuke},
  journal = {Phys. Rev. A},
  volume = {79},
  issue = {1},
  pages = {013627},
  numpages = {5},
  year = {2009},
  month = {Jan},
  publisher = {American Physical Society},
  doi = {10.1103/PhysRevA.79.013627},
  url = {https://link.aps.org/doi/10.1103/PhysRevA.79.013627}
}

@article{Chen2005PR,
  title = {{BCS–BEC} crossover: {From} high temperature superconductors to ultracold superfluids},
  journal = {Physics Reports},
  volume = {412},
  number = {1},
  pages = {1-88},
  year = {2005},
  issn = {0370-1573},
  doi = {https://doi.org/10.1016/j.physrep.2005.02.005},
  url = {https://www.sciencedirect.com/science/article/pii/S0370157305001067},
  author = {Qi Jin Chen and Jelena Stajic and Shina Tan and K. Levin}
}

@article{Timusk1999RPP,
  title = {The pseudogap in high-temperature superconductors: an experimental survey},
  author = {Tom Timusk and  Bryan Statt},
  year = {1999},
  month = {jan},
  publisher = {},
  volume = {62},
  number = {1},
  pages = {61},
  journal = {Reports on Progress in Physics},
  doi = {10.1088/0034-4885/62/1/002},
  url = {https://dx.doi.org/10.1088/0034-4885/62/1/002},
}

@article{Randeria2014ARCMP,
  author = {Randeria, Mohit and Taylor, Edward},
  title = {Crossover from Bardeen-Cooper-Schrieffer to Bose-Einstein Condensation and the Unitary Fermi Gas}, 
  journal = {Annual Review of Condensed Matter Physics},
  year = {2014},
  volume = {5},
  number = {Volume 5, 2014},
  pages = {209-232},
  doi = {https://doi.org/10.1146/annurev-conmatphys-031113-133829},
  url = {https://www.annualreviews.org/content/journals/10.1146/annurev-conmatphys-031113-133829},
  publisher = {Annual Reviews},
  issn = {1947-5462},
  type = {Journal Article},
}

@article{Bloch2012NP,
  Title = {Quantum simulations with ultracold quantum gases},
  Author = {Bloch, Immanuel and Dalibard, Jean and Nascimbene, Sylvain},
  Journal = {Nature Physics},
  Year = {2012},
  Volume = {8},
  Number = {4},
  Pages = {267-276},
  Month = {APR},
  DOI = {10.1038/NPHYS2259},
  ISSN = {1745-2473},
  EISSN = {1745-2481}
}

@article{Chin2010RMP,
  title = {Feshbach resonances in ultracold gases},
  author = {Chin, Cheng and Grimm, Rudolf and Julienne, Paul and Tiesinga, Eite},
  journal = {Rev. Mod. Phys.},
  volume = {82},
  issue = {2},
  pages = {1225--1286},
  numpages = {0},
  year = {2010},
  month = {Apr},
  publisher = {American Physical Society},
  doi = {10.1103/RevModPhys.82.1225},
  url = {https://link.aps.org/doi/10.1103/RevModPhys.82.1225}
}

@article{Vale2021NP,
  Title = {Spectroscopic probes of quantum gases},
  Author = {Vale, Chris J. and Zwierlein, Martin},
  Journal = {Nature Physics},
  Year = {2021},
  Volume = {17},
  Number = {12},
  Pages = {1305-1315},
  Month = {DEC},
  DOI = {10.1038/s41567-021-01434-6},
  ISSN = {1745-2473},
  EISSN = {1745-2481}
}

@article{Torma2016PC,
  title = {Physics of ultracold Fermi gases revealed by spectroscopies},
  author = {T\"orm\"a, P\"aivi},
  journal = {Physica Scripta},
  year = {2016},
  month = {mar},
  publisher = {IOP Publishing},
  volume = {91},
  number = {4},
  pages = {043006},
  doi = {10.1088/0031-8949/91/4/043006},
  url = {https://dx.doi.org/10.1088/0031-8949/91/4/043006}
}

@article{Schirotzek140403PRL,
  title = {Determination of the Superfluid Gap in Atomic Fermi Gases by Quasiparticle Spectroscopy},
  author = {Schirotzek, Andr\'e and Shin, Yong-il and Schunck, Christian H. and Ketterle, Wolfgang},
  journal = {Phys. Rev. Lett.},
  volume = {101},
  issue = {14},
  pages = {140403},
  numpages = {4},
  year = {2008},
  month = {Oct},
  publisher = {American Physical Society},
  doi = {10.1103/PhysRevLett.101.140403},
  url = {https://link.aps.org/doi/10.1103/PhysRevLett.101.140403}
}

@article{Stewart2010PRL,
  title = {Verification of Universal Relations in a Strongly Interacting {Fermi} Gas},
  author = {Stewart, J. T. and Gaebler, J. P. and Drake, T. E. and Jin, D. S.},
  journal = {Phys. Rev. Lett.},
  volume = {104},
  issue = {23},
  pages = {235301},
  numpages = {4},
  year = {2010},
  month = {Jun},
  publisher = {American Physical Society},
  doi = {10.1103/PhysRevLett.104.235301},
  url = {https://link.aps.org/doi/10.1103/PhysRevLett.104.235301}
}

@article{Sagi2012PRL,
  title = {Measurement of the Homogeneous Contact of a Unitary {Fermi} Gas},
  author = {Sagi, Yoav and Drake, Tara E. and Paudel, Rabin and Jin, Deborah S.},
  journal = {Phys. Rev. Lett.},
  volume = {109},
  issue = {22},
  pages = {220402},
  numpages = {4},
  year = {2012},
  month = {Nov},
  publisher = {American Physical Society},
  doi = {10.1103/PhysRevLett.109.220402},
  url = {https://link.aps.org/doi/10.1103/PhysRevLett.109.220402}
}

@article{Feld2011N,
  Title = {Observation of a pairing pseudogap in a two-dimensional {Fermi} gas},
  Author = {Feld, Michael and Froehlich, Bernd and Vogt, Enrico and Koschorreck, Marco and Koehl, Michael},
  Journal = {Nature},
  Year = {2011},
  Volume = {480},
  Number = {7375},
  Pages = {75-U233},
  Month = {DEC 1},
  DOI = {10.1038/nature10627},
  ISSN = {0028-0836},
  EISSN = {1476-4687}
}

@PhdThesis{ChenPhD,
  author = {Chen, Qi Jin},
  title = {Generalization of {BCS} theory to short coherence length superconductors: {A BCS-Bose-Einstein} crossover scenario},
  school = {University of Chicago},
  year = {2000},
  note = {available as arXiv:1801.06266.},
  url = {http://arxiv.org/abs/1801.06266}
}

@article{Chen1998PRL,
  title = {Pairing Fluctuation Theory of Superconducting Properties in Underdoped to Overdoped Cuprates},
  author = {Chen, Qi Jin and Kosztin, Ioan and Jank\'o, Boldizs\'ar and Levin, K.},
  journal = {Phys. Rev. Lett.},
  volume = {81},
  issue = {21},
  pages = {4708--4711},
  numpages = {0},
  year = {1998},
  month = {Nov},
  publisher = {American Physical Society},
  doi = {10.1103/PhysRevLett.81.4708},
  url = {https://link.aps.org/doi/10.1103/PhysRevLett.81.4708}
}

@article{Chen1999PRB,
  title = {Superconducting transitions from the pseudogap state: d-wave symmetry, lattice, and low-dimensional effects},
  author = {Chen, Qi Jin and Kosztin, Ioan and Jank\'o, Boldizs\'ar and Levin, K.},
  journal = {Phys. Rev. B},
  volume = {59},
  issue = {10},
  pages = {7083--7093},
  numpages = {0},
  year = {1999},
  month = {Mar},
  publisher = {American Physical Society},
  doi = {10.1103/PhysRevB.59.7083},
  url = {https://link.aps.org/doi/10.1103/PhysRevB.59.7083}
}

@misc{letter,
  title = {Spectral study of the pseudogap in unitary Fermi gases},
  author = {Chu Ping Li and Lin Sun and Kai Chao Zhang and Jun Ru Wu and Yu Xuan Wu and Ding Li Yuan and Peng Yi Chen and Qi Jin Chen},
  year = {2026}
}

@article{Johansen2024PRA,
  title = {Spectral functions of the strongly interacting three-dimensional {Fermi} gas},
  author = {Johansen, Christian H. and Frank, Bernhard and Lang, Johannes},
  journal = {Phys. Rev. A},
  volume = {109},
  issue = {2},
  pages = {023324},
  numpages = {19},
  year = {2024},
  month = {Feb},
  publisher = {American Physical Society},
  doi = {10.1103/PhysRevA.109.023324},
  url = {https://link.aps.org/doi/10.1103/PhysRevA.109.023324}
}

@article{Enss2024PRA,
  title = {Particle and pair spectra for strongly correlated {Fermi} gases: A real-frequency solver},
  author = {Enss, Tilman},
  journal = {Phys. Rev. A},
  volume = {109},
  issue = {2},
  pages = {023325},
  numpages = {11},
  year = {2024},
  month = {Feb},
  publisher = {American Physical Society},
  doi = {10.1103/PhysRevA.109.023325},
  url = {https://link.aps.org/doi/10.1103/PhysRevA.109.023325}
}

@Article{Yin2009,
  author = 	 {Zeng-Qiang Yu and Kun Huang and Lan Yin},
  title = 	 {Induced interaction in a {F}ermi gas with a {BEC-BCS} crossover},
  journal = 	 {\pra},
  year = 	 2009,
  volume = 	 79,
  pages = 	 {053636}
}

@Article{Li2024N,
  author={Li, Xi and Wang, Shuai and Luo, Xiang and Zhou, Yu-Yang and Xie, Ke and Shen, Hong-Chi and Nie, Yu-Zhao and Chen, Qi Jin and Hu, Hui and Chen, Yu-Ao and Yao, Xing-Can and Pan, Jian-Wei},
  title={Observation and quantification of the pseudogap in unitary {Fermi} gases},
  journal={Nature},
  year={2024},
  month={Feb},
  day={01},
  volume={626},
  number={7998},
  pages={288-293},
  issn={1476-4687},
  doi={10.1038/s41586-023-06964-y},
  url={https://doi.org/10.1038/s41586-023-06964-y}
}

@article{Chin2004S,
  title = {Observation of the pairing gap in a strongly interacting {Fermi} gas},
  Author = {Chin, C and Bartenstein, M and Altmeyer, A and Riedl, S and Jochim, S and Denschlag, JH and Grimm, R},
  Journal = {Science},
  Year = {2004},
  Volume = {305},
  Number = {5687},
  Pages = {1128-1130},
  Month = {AUG 20},
  DOI = {10.1126/science.1100818},
  ISSN = {0036-8075}
}

@article{Ding1996N,
  author={Ding, H. and Yokoya, T. and Campuzano, J. C. and Takahashi, T. and Randeria, M. and Norman, M. R. and Mochiku, T. and Kadowaki, K. and Giapintzakis, J.},
  title={Spectroscopic evidence for a pseudogap in the normal state of underdoped high-${T}_{c}$ superconductors},
  journal={Nature},
  year={1996},
  month={Jul},
  day={01},
  volume={382},
  number={6586},
  pages={51-54},
  issn={1476-4687},
  doi={10.1038/382051a0},
  url={https://doi.org/10.1038/382051a0}
}

@article{Chen2014FP,
  Author = {Chen, Qi Jin and Wang, Ji Biao},
  Title = {Pseudogap phenomena in ultracold atomic {Fermi} gases},
  Journal = {FRONTIERS OF PHYSICS},
  Year = {2014},
  Volume = {9},
  Number = {5},
  Pages = {539-570},
  Month = {OCT},
  DOI = {10.1007/s11467-014-0448-7},
  ISSN = {2095-0462},
  url={https://doi.org/10.1007/s11467-014-0448-7}
}

@article{Stewart2008N,
  Title = {Using photoemission spectroscopy to probe a strongly interacting {Fermi} gas},
  Author = {Stewart, J. T. and Gaebler, J. P. and Jin, D. S.},
  Journal = {Nature},
  Year = {2008},
  Volume = {454},
  Number = {7205},
  Pages = {744-747},
  Month = {AUG 7},
  DOI = {10.1038/nature07172},
  ISSN = {0028-0836}
}

@article{Gaebler2010NP,
  Author = {Gaebler, J. P. and Stewart, J. T. and Drake, T. E. and Jin, D. S. and
  Perali, A. and Pieri, P. and Strinati, G. C.},
  Title = {Observation of pseudogap behaviour in a strongly interacting {Fermi} gas},
  Journal = {Nature Physics},
  Year = {2010},
  Volume = {6},
  Number = {8},
  Pages = {569-573},
  Month = {AUG},
  DOI = {10.1038/NPHYS1709},
  ISSN = {1745-2473},
  EISSN = {1745-2481},
}

@article{Sagi2015PRL,
  title = {Breakdown of the {Fermi} Liquid Description for Strongly Interacting Fermions},
  author = {Sagi, Yoav and Drake, Tara E. and Paudel, Rabin and Chapurin, Roman and Jin, Deborah S.},
  journal = {Phys. Rev. Lett.},
  volume = {114},
  issue = {7},
  pages = {075301},
  numpages = {5},
  year = {2015},
  month = {Feb},
  publisher = {American Physical Society},
  doi = {10.1103/PhysRevLett.114.075301},
  url = {https://link.aps.org/doi/10.1103/PhysRevLett.114.075301}
}

@article{Perali2008PRL,
  title = {Competition between Final-State and Pairing-Gap Effects in the Radio-Frequency Spectra of Ultracold {Fermi} Atoms},
  author = {Perali, A. and Pieri, P. and Strinati, G. C.},
  journal = {Phys. Rev. Lett.},
  volume = {100},
  issue = {1},
  pages = {010402},
  numpages = {4},
  year = {2008},
  month = {Jan},
  publisher = {American Physical Society},
  doi = {10.1103/PhysRevLett.100.010402},
  url = {https://link.aps.org/doi/10.1103/PhysRevLett.100.010402}
}

@article{Magierski2009PRL,
  title = {Finite-Temperature Pairing Gap of a Unitary {Fermi} Gas by Quantum {Monte Carlo} Calculations},
  author = {Magierski, Piotr and Wlaz\l{}owski, Gabriel and Bulgac, Aurel and Drut, Joaqu\'{\i}n E.},
  journal = {Phys. Rev. Lett.},
  volume = {103},
  issue = {21},
  pages = {210403},
  numpages = {4},
  year = {2009},
  month = {Nov},
  publisher = {American Physical Society},
  doi = {10.1103/PhysRevLett.103.210403},
  url = {https://link.aps.org/doi/10.1103/PhysRevLett.103.210403}
}

@article{Chen2009PRL,
  title = {Momentum Resolved Radio Frequency Spectroscopy in Trapped {Fermi} Gases},
  author = {Chen, Qi Jin and Levin, K.},
  journal = {Phys. Rev. Lett.},
  volume = {102},
  issue = {19},
  pages = {190402},
  numpages = {4},
  year = {2009},
  month = {May},
  publisher = {American Physical Society},
  doi = {10.1103/PhysRevLett.102.190402},
  url = {https://link.aps.org/doi/10.1103/PhysRevLett.102.190402}
}

@article{Nishida2007PRA,
  title = {Fermi gas near unitarity around four and two spatial dimensions},
  author = {Nishida, Yusuke and Son, Dam Thanh},
  journal = {Phys. Rev. A},
  volume = {75},
  issue = {6},
  pages = {063617},
  numpages = {22},
  year = {2007},
  month = {Jun},
  publisher = {American Physical Society},
  doi = {10.1103/PhysRevA.75.063617},
  url = {https://link.aps.org/doi/10.1103/PhysRevA.75.063617}
}

@article{Haussmann2009PRA,
  title = {Spectral functions and rf response of ultracold fermionic atoms},
  author = {Haussmann, R. and Punk, M. and Zwerger, W.},
  journal = {Phys. Rev. A},
  volume = {80},
  issue = {6},
  pages = {063612},
  numpages = {18},
  year = {2009},
  month = {Dec},
  publisher = {American Physical Society},
  doi = {10.1103/PhysRevA.80.063612},
  url = {https://link.aps.org/doi/10.1103/PhysRevA.80.063612}
}

@article{Chen2016SR,
  author={Chen, Qi Jin},
  title={Effect of the particle-hole channel on {BCS--Bose-Einstein} condensation crossover in atomic {Fermi} gases},
  journal={Scientific Reports},
  year={2016},
  month={May},
  day={17},
  volume={6},
  number={1},
  pages={25772},
  issn={2045-2322},
  doi={10.1038/srep25772},
  url={https://doi.org/10.1038/srep25772}
}

@Article{Ku2012S,
  author = {Ku, Mark J. H. and Sommer, Ariel T. and Cheuk, Lawrence W. and Zwierlein, Martin W.},
  journal = {SCIENCE},
  title = {Revealing the Superfluid Lambda Transition in the Universal Thermodynamics of a Unitary Fermi Gas},
  year = {2012},
  issn = {0036-8075},
  month = {FEB 3},
  number = {6068},
  pages = {563-567},
  volume = {335},
  doi = {10.1126/science.1214987},
  eissn = {1095-9203}
}

@article{Biss2022PRL,
  title = {Excitation Spectrum and Superfluid Gap of an Ultracold {Fermi} Gas},
  author = {Biss, Hauke and Sobirey, Lennart and Luick, Niclas and Bohlen, Markus and Kinnunen, Jami J. and Bruun, Georg M. and Lompe, Thomas and Moritz, Henning},
  journal = {Phys. Rev. Lett.},
  volume = {128},
  issue = {10},
  pages = {100401},
  numpages = {7},
  year = {2022},
  month = {Mar},
  publisher = {American Physical Society},
  doi = {10.1103/PhysRevLett.128.100401},
  url = {https://link.aps.org/doi/10.1103/PhysRevLett.128.100401}
}

@article{Mukherjee2019PRL,
  title = {Spectral Response and Contact of the Unitary {Fermi} Gas},
  author = {Mukherjee, Biswaroop and Patel, Parth B. and Yan, Zhenjie and Fletcher, Richard J. and Struck, Julian and Zwierlein, Martin W.},
  journal = {Phys. Rev. Lett.},
  volume = {122},
  issue = {20},
  pages = {203402},
  numpages = {6},
  year = {2019},
  month = {May},
  publisher = {American Physical Society},
  doi = {10.1103/PhysRevLett.122.203402},
  url = {https://link.aps.org/doi/10.1103/PhysRevLett.122.203402}
}

@article{Carlson2011PRA,
  title = {Auxiliary-field quantum {Monte Carlo} method for strongly paired fermions},
  author = {Carlson, J. and Gandolfi, Stefano and Schmidt, Kevin E. and Zhang, Shiwei},
  journal = {Phys. Rev. A},
  volume = {84},
  issue = {6},
  pages = {061602},
  numpages = {5},
  year = {2011},
  month = {Dec},
  publisher = {American Physical Society},
  doi = {10.1103/PhysRevA.84.061602},
  url = {https://link.aps.org/doi/10.1103/PhysRevA.84.061602}
}

@article{Endres2013PRA,
  title = {Lattice {Monte Carlo} calculations for unitary fermions in a finite box},
  author = {Endres, Michael G. and Kaplan, David B. and Lee, Jong-Wan and Nicholson, Amy N.},
  journal = {Phys. Rev. A},
  volume = {87},
  issue = {2},
  pages = {023615},
  numpages = {17},
  year = {2013},
  month = {Feb},
  publisher = {American Physical Society},
  doi = {10.1103/PhysRevA.87.023615},
  url = {https://link.aps.org/doi/10.1103/PhysRevA.87.023615}
}

@article{Pessoa2015PRA,
  title = {Contact interaction in a unitary ultracold {Fermi} gas},
  author = {Pessoa, Renato and Gandolfi, S. and Vitiello, S. A. and Schmidt, K. E.},
  journal = {Phys. Rev. A},
  volume = {92},
  issue = {6},
  pages = {063625},
  numpages = {7},
  year = {2015},
  month = {Dec},
  publisher = {American Physical Society},
  doi = {10.1103/PhysRevA.92.063625},
  url = {https://link.aps.org/doi/10.1103/PhysRevA.92.063625}
}

@article{Zurn2013PRL,
  title = {Precise Characterization of $^{6}\mathrm{Li}$ {Feshbach} Resonances Using Trap-Sideband-Resolved RF Spectroscopy of Weakly Bound Molecules},
  author = {Z\"urn, G. and Lompe, T. and Wenz, A. N. and Jochim, S. and Julienne, P. S. and Hutson, J. M.},
  journal = {Phys. Rev. Lett.},
  volume = {110},
  issue = {13},
  pages = {135301},
  numpages = {5},
  year = {2013},
  month = {Mar},
  publisher = {American Physical Society},
  doi = {10.1103/PhysRevLett.110.135301},
  url = {https://link.aps.org/doi/10.1103/PhysRevLett.110.135301}
}

@article{Reichl2015PRA,
  title = {Quasiparticle dispersions and lifetimes in the normal state of the {BCS-BEC} crossover},
  author = {Reichl, Matthew D. and Mueller, Erich J.},
  journal = {Phys. Rev. A},
  volume = {91},
  issue = {4},
  pages = {043627},
  numpages = {5},
  year = {2015},
  month = {Apr},
  publisher = {American Physical Society},
  doi = {10.1103/PhysRevA.91.043627},
  url = {https://link.aps.org/doi/10.1103/PhysRevA.91.043627}
}

@article{GORKOV1961,
  title={Contribution to the theory of superfluidity in an imperfect {Fermi} gas},
  author={Gor’kov, LP and Melik-Barkhudarov, TK},
  journal={Sov. Phys. JETP},
  volume={13},
  number={5},
  Pages = {1018-1022},
  year={1961}
}

@article{Chen2009RoPP,
  doi = {10.1088/0034-4885/72/12/122501},
  url = {https://dx.doi.org/10.1088/0034-4885/72/12/122501},
  year = {2009},
  month = {oct},
  publisher = {},
  volume = {72},
  number = {12},
  pages = {122501},
  author = {Qi Jin Chen and Yan He and Chih-Chun Chien and K Levin},
  title = {Theory of radio frequency spectroscopy experiments in ultracold Fermi gases and their relation to photoemission in the cuprates},
  journal = {Reports on Progress in Physics}
}

@article{DDW,
   author = {Sudip Chakravarty and R. B. Laughlin and Dirk K. Morr and Chetan Nayak},
   doi = {10.1103/PhysRevB.63.094503},
   issn = {0163-1829},
   issue = {9},
   journal = {Physical Review B},
   month = {1},
   pages = {094503},
   title = {Hidden order in the cuprates},
   volume = {63},
   url = {https://link.aps.org/doi/10.1103/PhysRevB.63.094503},
   year = {2001}
}

@article{Varma2014,
doi = {10.1088/0953-8984/26/50/505701},
url = {https://doi.org/10.1088/0953-8984/26/50/505701},
year = {2014},
month = {nov},
publisher = {IOP Publishing},
volume = {26},
number = {50},
pages = {505701},
author = {Varma, C M},
title = {Pseudogap in cuprates in the loop-current ordered state},
journal = {Journal of Physics: Condensed Matter}
}

@InProceedings{Leggett1980,
  author    = {Leggett, Anthony James},
  booktitle = {Modern trends in the theory of condensed matter},
  title     = {Diatomic molecules and {Cooper} pairs},
  year      = {1980},
  address   = {Berlin, West Germany},
  editor    = {Andrzej Pekalski and Jerzy A. Przystawa},
  note      = {Proceedings of the XVI Karpacz Winter School of Theoretical Physics, February 19 - March 3, 1979, Karpacz, Poland},
  pages     = {13--27},
  publisher = {Springer-Verlag},
  series    = {Lecture Notes in Physics},
  volume    = {115},
  url       = {https://link.springer.com/content/pdf/10.1007/BFb0120125},
}

@phdthesis{Regal2005,
   author = {Cindy Regal},
   issue = {December},
   school = {Univ Colorado, Boulder},
   pages = {1-149},
   title = {Experimental realization of BCS-BEC crossover physics with a Fermi gas of atoms},
   year = {2005}
}

@Article{Kadanoff1961,
  author    = {Kadanoff, Leo P. and Martin, Paul C.},
  journal   = {Phys. Rev.},
  title     = {Theory of Many-Particle Systems. {II. Superconductivity}},
  year      = {1961},
  month     = {Nov},
  pages     = {670--697},
  volume    = {124},
  doi       = {10.1103/PhysRev.124.670},
  issue     = {3},
  numpages  = {0},
  publisher = {American Physical Society},
  url       = {https://link.aps.org/doi/10.1103/PhysRev.124.670},
}

@article{Dizer,
  title = {Spectral properties and observables in ultracold Fermi gases},
  author = {Dizer, Eugen and Horak, Jan and Pawlowski, Jan M.},
  journal = {Phys. Rev. A},
  volume = {109},
  issue = {6},
  pages = {063311},
  numpages = {17},
  year = {2024},
  month = {Jun},
  publisher = {American Physical Society},
  doi = {10.1103/PhysRevA.109.063311},
  url = {https://link.aps.org/doi/10.1103/PhysRevA.109.063311}
}

@InCollection{Strinati2012,
  author = 	 {G. C. Strinati},
  title = 	 {Pairing Fluctuations Approach to the {BCS–BEC} Crossover},
  booktitle = 	 {The {BCS-BEC} crossover and the unitary {Fermi} gas},
  publisher = {Springer-Verlag},
  address = 	 {Berlin, Heidelberg},
  year = 	 2012,
  editor = 	 {Zwerger, Wilhelm},
  volume = 	 836,
  series = 	 {Lecture Notes in Physics},
  chapter = 	 4,
  pages = 	 {99--126},
  doi       = {10.1007/978-3-642-21978-8_4},
}

@article{Tchernyshyov1997,
   author = {Oleg Tchernyshyov},
   doi = {10.1103/PhysRevB.56.3372},
   issn = {0163-1829},
   issue = {6},
   journal = {Physical Review B},
   month = {8},
   pages = {3372-3380},
   title = {Noninteracting Cooper pairs inside a pseudogap},
   volume = {56},
   url = {https://link.aps.org/doi/10.1103/PhysRevB.56.3372},
   year = {1997}
}

@article{Johansen2024,
  title = {Spectral functions of the strongly interacting three-dimensional Fermi gas},
  author = {Johansen, Christian H. and Frank, Bernhard and Lang, Johannes},
  journal = {Phys. Rev. A},
  volume = {109},
  issue = {2},
  pages = {023324},
  numpages = {19},
  year = {2024},
  month = {Feb},
  publisher = {American Physical Society},
  doi = {10.1103/PhysRevA.109.023324},
  url = {https://link.aps.org/doi/10.1103/PhysRevA.109.023324}
}

@article{Enss2024,
  title = {Particle and pair spectra for strongly correlated Fermi gases: A real-frequency solver},
  author = {Enss, Tilman},
  journal = {Phys. Rev. A},
  volume = {109},
  issue = {2},
  pages = {023325},
  numpages = {11},
  year = {2024},
  month = {Feb},
  publisher = {American Physical Society},
  doi = {10.1103/PhysRevA.109.023325},
  url = {https://link.aps.org/doi/10.1103/PhysRevA.109.023325}
}

\end{document}